\documentclass{aa} 
\usepackage{graphicx}
\usepackage{txfonts}
\usepackage{natbib}
\usepackage{verbatim}
\usepackage{longtable}
\usepackage{afterpage}
\bibpunct{(}{)}{;}{a}{}{,}
\begin{document}

\title{Estimated IR and phosphorescence emission fluxes for 
specific Polycyclic Aromatic Hydrocarbons in the Red Rectangle}

\author{G. Mulas\inst{1}
	\and 
	G. Malloci\inst{1}
	\and
	C. Joblin\inst{2}
	\and 
	D. Toublanc\inst{2}
}

\offprints{G.~Mulas,\\\email{gmulas@ca.astro.it}}

\institute{INAF~\textendash~Osservatorio Astronomico di 
Cagliari~\textendash~Astrochemistry Group, 
Strada n.54, Loc. Poggio dei Pini, I\textendash09012 Capoterra  (CA), Italy\\
\email{[gmulas; gmalloci]@ca.astro.it}
\and
Centre d'Etude Spatiale des Rayonnements, CNRS et Universit\'e Paul Sabatier,
Observatoire Midi-Pyr\'en\'ees, 9 Avenue du colonel Roche, 
31028 Toulouse cedex 04, France\\
\email{[christine.joblin; dominique.toublanc]@cesr.fr}
}

\date{Received July 1 2005/ Accepted September 14 2005}

\abstract{Following the tentative identification of the blue luminescence
in the Red Rectangle by \citet{vij05}, we compute absolute fluxes 
for the vibrational IR emission and phosphorescence bands of three small 
polycyclic aromatic hydrocarbons. The calculated IR spectra are compared with 
available ISO observations. A subset of the emission bands are predicted to be 
observable using presently available facilities, and can be used for an 
immediate, independent, discriminating test on their alleged presence in this 
well\textendash known astronomical object.

}

\authorrunning{Mulas et al.}
\titlerunning{Absolute IR and phosphorescence fluxes from PAHs 
in the Red Rectangle}

\maketitle

\section{Introduction}\label{introduction}
Recently, \citet{vij04} detected a blue luminescence (BL) in the well\textendash known 
bipolar protoplanetary Red Rectangle (RR) nebula \citep{coh75,coh04}.
Shortly thereafter, \citet{vij05}, after a thorough analysis of
the extinction properties and of the correlation of the BL with the 
observed near\textendash IR emission at 3.3~$\mu$m, 
concluded that they are both most likely due to small, neutral polycyclic
aromatic hydrocarbons (PAHs), composed of three to four aromatic units.
Another possible explanation of the BL tentatively identifies it as 
fluorescence by ultrasmall silicon nanoparticles \citep{nay05}.

The hypothesis of the ubiquitous presence of free gas\textendash phase PAHs in the 
interstellar medium (ISM) originated about 20 years ago \citep{leg84,all85}. 
Due to their spectral properties and their high photostability, these 
molecules were suggested as the most natural interpretation for the 
so\textendash called ``Aromatic Infrared Bands'' (AIBs), a set of emission bands 
observed near 3.3, 6.2, 7.7, 8.6, 11.3 and 12.7~$\mu$m, in many dusty 
environments excited by UV photons \citep{leg89,all89}. The AIBs are the 
spectral fingerprint of the excitation of vibrations in aromatic C\textendash C and 
C\textendash H bonds \citep{dul81}. Furthermore, PAHs and their cations were also 
supposed to account for a subset of the ``Diffuse Interstellar Bands'' 
(DIBs) \citep{leg85,van85,cra85}, more than $\sim300$ absorption features 
ubiquitously observed in the near-UV, visible and near-IR in the spectra 
of reddened stars \citep[see e.~g.][ and references therein]{ehr00}. 

The RR nebula, besides being one of the brightest sources of AIBs
in the sky, also has the almost unique property of displaying a subset of 
emission bands in the visible \citep{sch80,van02} which appear to be 
associated with a subset of DIBs \citep{sca92}. The RR and the R~CrB star 
V854 Cen \citep{rao93} are the \emph{only} known cases in which DIBs have 
ever been possibly observed in emission. This makes the tentative 
identification of small neutral PAHs in this same nebula a potentially 
very exciting discovery, with far\textendash reaching consequences also on the 
long\textendash standing mystery of the identification of DIBs.

However, while the analysis in \citet{vij05} narrows the range of 
potential carriers of the observed BL to a small
number of molecules, it cannot provide a definitive identification, 
which requires an independent test. Such a test can be obtained from
the IR and phosphorescence emission spectrum of the same 
molecules. The low\textendash frequency vibrational modes 
\citep{lan96,zha96} and phosphorescence bands \citep{sal04} provide an unique 
signature and ought to be produced together with electronic fluorescence bands
\citep{bre05}.

We carried out simulations of the photophysics of the candidate
molecules using a Monte\textendash Carlo code 
\citep{mul98,job02,mal03c,mul03}, together with quantum\textendash 
chemical calculations for the relevant molecular 
parameters \citep{lan96,mar96,mal04} and available laboratory measurements 
for the photoabsorption spectra \citep{job92b,job92a} and for visible and IR 
fluorescence quantum yields \citep{bre05}.
This produces a quantitative prediction of the IR and 
phosphorescence emission spectra for each given molecule, which must be 
related to the integrated BL attributed to this same molecule.

\section{Modelling procedure}

In previous papers \citep{job02,mal03c,mul03} we demonstrated the use of 
Monte\textendash Carlo models to simulate the detailed photophysics of a specific PAH
embedded in a given radiation field (RF). We here apply the same 
procedure to the specific case of the three small, neutral PAHs which 
were tentatively identified in the Red Rectangle nebula by \citet{vij05}, 
namely anthracene (C$_{14}$H$_{10}$), phenanthrene (C$_{14}$H$_{10}$) and pyrene 
(C$_{16}$H$_{10}$) to derive their expected complete IR emission spectra.

For the modelling procedure we used quantum\textendash chemical results for structural 
parameters and vibrational analysis, both from the literature 
\citep{lan96,mar96} and obtained by ourselves. For anthracene and 
pyrene we used the experimental photo\textendash absorption cross\textendash sections from 
\citet{job92b,job92a}, while for phenanthrene we used the theoretical spectrum 
from \citet{mal04}. 
We performed all of the quantum\textendash chemical calculations in the framework 
of the Density Functional Theory (DFT), using the \textsc{Octopus} and 
\textsc{NWChem} computational codes. Details on them 
can be found elsewhere \citep{mal04,mal05}. 

A crucial modelling parameter is the assumed knowledge of the relaxation 
paths of the modelled molecules upon electronic excitation following a 
photon absorption. Almost all neutral PAHs, including
in particular anthracene, phenanthrene and pyrene, upon excitation 
\mbox{$\mathrm{S}_n\gets\mathrm{S}_0$}, are well\textendash known to 
undergo very fast internal conversion (IC) to the $\mathrm{S}_1$ electronic 
level \citep[see e.~g.][ and references therein]{lea95}. From there, three 
relaxation paths are available, their relative 
weights dependent on the vibrational energy available: \begin{enumerate}
\item fluorescence \mbox{$\mathrm{S}_1\to\mathrm{S}_0$} with a permitted 
electronic transition, the remaining energy being subsequently radiated 
by vibrational transitions in $\mathrm{S}_0$; relaxation of small PAHs via 
this path is the proposed origin of BL in the RR \citep{vij04,vij05};
\item direct \mbox{$\mathrm{S}_1\leadsto\mathrm{T}_1$} or indirect 
\mbox{$\mathrm{S}_1\leadsto\mathrm{T}_n\leadsto\mathrm{T}_1$} intersystem crossing 
(ISC), a radiationless transition followed by the emission of a 
phosphorescence photon in a \mbox{$\mathrm{T}_1\to\mathrm{S}_0$} spin\textendash forbidden, 
electronic\textendash permitted transition; the remaining energy is radiated in 
vibrational transitions either (almost always) from $\mathrm{S}_0$ after 
the phosphorescence transition or (very seldom) from $\mathrm{T}_1$ before it;
\item internal conversion \mbox{$\mathrm{S}_1\leadsto\mathrm{S}_0$}, a radiationless 
transition, after which essentially all excitation energy is radiated by 
vibrational transitions.
\end{enumerate}
According to experimental results \citep{bre05}, the rate of fluorescence 
transitions (1) is essentially independent of the excitation energy. The 
rate of ISC radiationless transitions (2) increases slightly with excitation 
energy for the three molecules considered. The relaxation path (3) by IC to 
the ground state is open only when some excess vibrational energy in 
$\mathrm{S}_1$ is available, such threshold depending on the specific molecule 
and varying from $\sim2\,\,10^3$~cm$^{-1}$ for anthracene to $\sim4\,\,10^4$~cm$^{-1}$ 
for pyrene. Above this threshold the rate of IC to $\mathrm{S}_0$ (3) 
exponentially increases, becoming by and large the dominant relaxation path. 
The energy\textendash dependent quantum yields for these three relaxation paths were 
measured in gas\textendash phase experiments by \citet{bre05} for all the three 
molecules under study here. 

All of the three relaxation paths described above contribute to IR emission 
from the ground  $\mathrm{S}_0$ electronic state and are therefore considered
in our model. Even the slowest electronic 
transition, i.~e. the \mbox{$\mathrm{T}_1\to\mathrm{S}_0$} involved in 
phosphorescence, occurs with decay constants of the order of, at most, 
$\tau_\mathrm{ph}\sim1.5\,\,10^{-3}$~s \citep{sal04}, while the time constant involved 
in the fastest vibrational transitions is always $\tau_\mathrm{IR}\gtrsim5\,\,10^{-2}$~s. 
This ensures that vibrational relaxation almost always occurs \emph{after} 
any electronic transition in the relaxation path, i.~e. from 
$\mathrm{S}_0$. Along any of the three relaxation paths described above,
the only process which can to some extent compete with the IR emission is
the absorption of another UV\textendash visible photon before the end of the 
vibrational cascade. When this happens, the residual excitation energy is 
added to that of the newly absorbed photon, making available higher\textendash energy
vibrational modes. Emission in the weakest IR bands, mostly towards the 
low\textendash energy end, essentially happens only when the stronger ones, 
mostly towards the high\textendash energy end, are energetically unaccessible;
if vibrational cascades are interrupted before their end, weak bands are 
consequently suppressed, in favour of the stronger ones, which are emitted 
much more quickly when the molecule excitation energy is high enough. 

We considered the modelled molecules in two different regions of the 
nebula and, therefore, our simulations correspond to two different exciting 
RFs. The first one is a Kurucz spectrum with effective 
temperature T~=~8250~K, surface gravity $\log g = 1.5$ and total 
luminosity of 6050~L$_\odot$ \citep{vij05}, summed with a blackbody at 
T~=~60000~K to represent the He white dwarf companion with a total luminosity 
of 100~L$_\odot$ \citep{men02}, both truncated at the Lyman limit and diluted 
according to the geometry of the source \citep{men02} at given angular 
distances along the polar axes of the nebula. Extinction along the polar axes 
of the nebula is low and particularly gray \citep{men02,vij05} and we neglect 
it altogether. We will therefore assume the \emph{spectrum} of the 
estimated RF to be constant along the bipolar lobes, and just scale with a 
wavelength\textendash independent factor with distance, in the wavelength range 
which pumps the BL. 

The second RF we considered is the ``attenuated'' one, as shown in 
\citet{vij05}, corresponding to what is seen by a molecule in the 
halo of the RR nebula, out of the bipolar cones and out of the dense,
optically thick torus\textendash like dust shell surrounding the central binary
star. This RF is essentially due to heavily obscured light leaking 
through the torus and to light initially directed along the bipolar 
lobes and subsequently scattered into the halo. Since the halo itself 
is optically very thin \citep{men02,vij05}, we again assume the 
\emph{spectrum} of this RF to be constant and just scale with a 
wavelength\textendash independent factor with distance, over the energy range 
which pumps the BL.

Both RFs are limited on the high\textendash energy side by the Lyman limit at 13.6~eV, 
since higher energy photons are completely absorbed in a small 
H$_\mathrm{II}$ region close to the central evolved star \citep{men02}. 
Photons of energies lower than the absorption edge of the first permitted 
electronic transition, which for small, neutral PAHs is always higher than 
$\sim$2~eV, are irrelevant for our purposes. The branching ratios for the three
main relaxation channels following UV/visible photon absorption by 
anthracene, phenanthrene and pyrene in the two RFs considered are listed
in Table~\ref{branching}. For the sake of completeness, we also included 
the ionisation yields, estimated according to the formula given by 
\citet{lep01}, using the ionisation potentials 7.439~eV for 
anthracene, 7.891~eV for phenanthrene and 7.426~eV for 
pyrene taken from the online NIST chemistry WebBook \citep{lia05}.

\begin{table}
\caption{Computed branching ratios for the three main relaxation channels
following UV/visible photon absorption by anthracene, phenanthrene and pyrene
in the RF of the lobes and halo of the RR nebula (see text for details).}
\label{branching}
\begin{center}
\begin{tabular}{lcccc}
\hline 
\hline
\noalign{\smallskip}
\noalign{\smallskip}
& \multicolumn{4}{c}{Relaxation branching ratios} \\
\noalign{\smallskip}
& fluorescence 
& I.~S.~C. 
& I.~C.
& ionisation \\
& $(\mathrm{S}_1\to\mathrm{S}_0)$
& $(\mathrm{S}_1\leadsto\mathrm{T}_1)$
& $(\mathrm{S}_1\leadsto\mathrm{S}_0)$
& \\
\noalign{\smallskip}
\cline{2-5}
\noalign{\smallskip}
\multicolumn{5}{c}{Anthracene (C$_{14}$H$_{10}$)}\\
\noalign{\smallskip}
lobes 
      & 0.157 & $\ll10^{-3}$ & 0.827 & 0.016 \\
halo  & 0.203 & $\ll10^{-3}$ & 0.796 & 0.001 \\
\noalign{\smallskip}
\cline{2-5}
\noalign{\smallskip}
\multicolumn{5}{c}{Phenanthrene (C$_{14}$H$_{10}$)}\\
\noalign{\smallskip}
lobes 
       & 0.031 & 0.372 & 0.587 & 0.010 \\
halo  & 0.042 & 0.562 & 0.396 & $3\,\,10^{-5}$ \\
\noalign{\smallskip}
\cline{2-5}
\noalign{\smallskip}
\multicolumn{5}{c}{Pyrene (C$_{16}$H$_{10}$)}\\
\noalign{\smallskip}
lobes
       & 0.060 & 0.284 & 0.642 & 0.014 \\
halo  & 0.094 & 0.408 & 0.497 & 0.001 \\
\noalign{\smallskip}
\hline
\hline
\end{tabular}
\end{center}
\end{table}

Under these conditions, the number 
$\displaystyle \frac{d R_\mathrm{bl}(\nu')}{d\nu'}$ of BL photons 
radiated per unit time by a given molecule in the frequency interval between 
$\nu'$ and $\nu'+d\nu'$ will be proportional to the respective fluorescence quantum 
yield $\displaystyle \frac{d Q_\mathrm{fl}(\nu,\nu')}{d\nu'}$, multiplied by the rate 
of absorption of exciting photons, in turn given by the product of 
the flux of exciting photons $F_\mathrm{exc}(\nu)$ times the absorption 
cross\textendash section $\sigma(\nu)$ of the molecule, i.~e.
\begin{eqnarray}
\frac{d R_\mathrm{bl}(\nu')}{d\nu'} & = & 
\int d\nu F_\mathrm{exc}(\nu)\sigma(\nu)\frac{d Q_\mathrm{fl}(\nu,\nu')}{d\nu'} \\
& = & \overline{F}_\mathrm{exc} 
\int d\nu \Phi_\mathrm{exc}(\nu)\sigma(\nu)\frac{d Q_\mathrm{fl}(\nu,\nu')}{d\nu'} \nonumber \\
& = & \nonumber \overline{F}_\mathrm{exc} \frac{d\overline{\sigma}_\mathrm{fl}(\nu')}
{d\nu'},
\end{eqnarray}
where we defined
\begin{displaymath}
\overline{F}_\mathrm{exc} = \int d\nu F_\mathrm{exc}(\nu), \qquad
\Phi_\mathrm{exc}(\nu) = \frac{F_\mathrm{exc}(\nu)}{\overline{F}_\mathrm{exc}}
\end{displaymath}
\begin{displaymath}
\mathrm{and}\quad \frac{d \overline{\sigma}_\mathrm{fl}(\nu')}{d\nu'} = 
\int d\nu \Phi_\mathrm{exc}(\nu)\sigma(\nu)\frac{d Q_\mathrm{fl}(\nu,\nu')}{d\nu'}.
\end{displaymath}
With the assumptions we made for the RFs, $\overline{F}_\mathrm{exc}$ will vary 
with the position in the nebula, while $\Phi_\mathrm{exc}(\nu)$, which contains the
frequency dependence of $F_\mathrm{exc}(\nu)$, will only vary between the 
biconic lobes and the halo but will be position\textendash independent within each 
of them. Since $\displaystyle \frac{d Q_\mathrm{fl}(\nu,\nu')}{d\nu'}$ is independent 
of the photon absorption rate, 
$\displaystyle \frac{d \overline{\sigma}_\mathrm{fl}(\nu')}{d\nu'}$ does not depend
on the position along the line of sight.

Completely analogous equations can be written for the IR emission and 
phosphorescence, i.~e.
\begin{eqnarray}
\frac{d R_\mathrm{IR}(\nu')}{d\nu'} & = & \overline{F}_\mathrm{exc} 
\int d\nu \Phi_\mathrm{exc}(\nu)\sigma(\nu)\frac{d Q_\mathrm{IR}(\nu,\nu')}{d\nu'} \\
& = & \nonumber \overline{F}_\mathrm{exc} 
\frac{d \overline{\sigma}_\mathrm{IR}(\nu')}{d\nu'}
\end{eqnarray}
\begin{displaymath}
\mathrm{with}\quad \frac{d \overline{\sigma}_\mathrm{IR}(\nu')}{d\nu'} = 
\int d\nu \Phi_\mathrm{exc}(\nu)\sigma(\nu)\frac{d Q_\mathrm{IR}(\nu,\nu')}{d\nu'}
\end{displaymath}
and
\begin{eqnarray}
\frac{{d R}_\mathrm{{ph}}{(\nu')}}{{d\nu'}} & {=} & {\overline{F}}_\mathrm{{exc}} 
{\int d\nu \Phi}_\mathrm{{exc}}{(\nu)\sigma(\nu)}\frac{{d Q}_\mathrm{{ph}}{(\nu,\nu')}}{{d\nu'}} \\
& {=} & \nonumber {\overline{F}}_\mathrm{{exc}} 
\frac{{d \overline{\sigma}}_\mathrm{{ph}}{(\nu')}}{{d\nu'}}
\end{eqnarray}
\begin{displaymath}
\mathrm{{with}}\quad \frac{{d \overline{\sigma}}_\mathrm{{ph}}{(\nu')}}{{d\nu'}} {= 
\int d\nu \Phi}_\mathrm{{exc}}{(\nu)\sigma(\nu)}\frac{{d Q}_\mathrm{{ph}}{(\nu,\nu')}}{{d\nu'}}.
\end{displaymath}

As explained above, $\displaystyle \frac{d Q_\mathrm{IR}(\nu,\nu')}{d\nu'}$
does slightly depend on the photon absorption rate if the average
time between photon absorptions is smaller than the average time it takes
for an excited molecule to complete its IR emission cascade. 
This is not the case for $\displaystyle \frac{d Q_\mathrm{ph}(\nu,\nu')}{d\nu'}$, 
since phosphorescence, although slower than fluorescence, still occurs on 
a much shorter timescale than the average interval between two photon 
absorptions in all cases considered here. The detailed three\textendash dimensional 
distribution of the molecules 
supposedly emitting the BL along a given line of sight is unknown:
the emitting molecules might equally well be concentrated close
to the source, where their photon absorption rate would be higher, or in 
a large volume extending relatively far out, with correspondingly lower
photon absorption rates. We therefore considered a range of distances to 
the central source, and hence of photon absorption rates, evaluating 
the corresponding variation of 
$\displaystyle \frac{d Q_\mathrm{IR}(\nu,\nu')}{d\nu'}$ and of the
calculated IR emission spectra.

Under optically thin conditions the observed fluorescence 
photon flux per unit solid angle on the sky along a given direction is 
given by:
\begin{eqnarray}
\frac{dF_\mathrm{bl}(\nu')}{d\Omega} & = & \label{bl_brightness}
\frac{1}{4\pi}\int dr n(r) \frac{d R_\mathrm{bl}(r,\nu')}{d\nu'} \\
& = & \frac{1}{4\pi} \frac{d \overline{\sigma}_\mathrm{fl}(\nu')}{d\nu'}
\int dr n(r) \overline{F}_\mathrm{exc}(r) \nonumber \\
& = & \nonumber \frac{1}{4\pi} \frac{d \overline{\sigma}_\mathrm{fl}(\nu')}{d\nu'} 
\Upsilon, 
\end{eqnarray}
with the definition
\begin{equation}
\label{upsilon_def}
\Upsilon = \int dr n(r) \overline{F}_\mathrm{exc}(r).
\end{equation}

Equation~(\ref{bl_brightness}) assumes 
$\displaystyle \frac{d \overline{\sigma}_\mathrm{fl}(\nu')}{d\nu'}$ to 
be constant along the line of sight, which is appropriate for a line of 
sight traversing only the halo of the RR. For a line of sight going 
through both the halo and one of the biconical lobes, the integral over the 
line of sight can be split in a part in the lobe and a part in the
halo, in each of which 
$\displaystyle \frac{d \overline{\sigma}_\mathrm{fl}(\nu')}{d\nu'}$ is constant.
Equation~(\ref{bl_brightness}) can be thus generalised to 
\begin{equation} \label{bl_brightness_2}
\frac{dF_\mathrm{bl}(\nu')}{d\Omega} = \frac{1}{4\pi} \left(
\frac{d \overline{\sigma}_\mathrm{fl}^\mathrm{lobe}(\nu')}{d\nu'} \Upsilon_\mathrm{lobe} +
\frac{d \overline{\sigma}_\mathrm{fl}^\mathrm{halo}(\nu')}{d\nu'} \Upsilon_\mathrm{halo}\right),
\end{equation}
where $\Upsilon_\mathrm{lobe}$ and $\Upsilon_\mathrm{halo}$ are still defined by 
Eq.~(\ref{upsilon_def}), the only difference being the domain of 
integration, which includes respectively only the part of the line 
of sight in the lobe or only the part in the halo.

Equivalent equations can be written again for both the IR emission and 
phosphorescence: 
\begin{equation} \label{IR_brightness_2}
\frac{dF_\mathrm{IR}(\nu')}{d\Omega} = \frac{1}{4\pi} \left(
\frac{d \overline{\sigma}_\mathrm{IR}^\mathrm{lobe}(\nu')}{d\nu'} \Upsilon_\mathrm{lobe}+
\frac{d \overline{\sigma}_\mathrm{IR}^\mathrm{halo}(\nu')}{d\nu'} \Upsilon_\mathrm{halo}\right).
\end{equation}
\begin{equation} \label{ph_brightness_2}
\frac{{dF}_\mathrm{{ph}}{(\nu')}}{{d\Omega}} 
= \frac{{1}}{{4\pi}} {\left( 
\frac{{d \overline{\sigma}}_\mathrm{{ph}}^\mathrm{{lobe}}{(\nu')}}{{d\nu'}} {\Upsilon}_\mathrm{{lobe}}+
\frac{{d \overline{\sigma}}_\mathrm{{ph}}^\mathrm{{halo}}{(\nu')}}{{d\nu'}}
{\Upsilon}_\mathrm{{halo}}\right)}.
\end{equation}
In Eq.~(\ref{IR_brightness_2}) above
$\displaystyle \frac{d \overline{\sigma}_\mathrm{IR}^\mathrm{lobe}(\nu')}{d\nu'}$
and $\displaystyle \frac{d \overline{\sigma}_\mathrm{IR}^\mathrm{halo}(\nu')}{d\nu'}$
are supposed to be effective values, averaged along the line of sight.

The quantities 
$\displaystyle \frac{d \overline{\sigma}_\mathrm{fl}^\mathrm{lobe}(\nu')}{d\nu'}$ 
and $\displaystyle \frac{d \overline{\sigma}_\mathrm{fl}^\mathrm{halo}(\nu')}{d\nu'}$ 
can be estimated for anthracene, pyrene and phenanthrene 
using the assumed spectrum of the exciting radiation fields and the 
quantum yields measured by \citet{bre05}. 
If we consider a line of sight dominated either by the lobes or by the 
halo, we can use the simpler Eq.~(\ref{bl_brightness}). Integrating both 
sides of Eq.~(\ref{bl_brightness}) over $\nu'$ we obtain
\begin{eqnarray} \label{bl_brightness_3}
\int d\nu' \frac{dF_\mathrm{bl}(\nu')}{d\Omega} & = & 
\frac{\Upsilon}{4\pi} \int d\nu' \frac{d \overline{\sigma}_\mathrm{fl}(\nu')}{d\nu'} \\
& = & \nonumber \frac{\Upsilon}{4\pi} 
\int d\nu \Phi_\mathrm{exc}(\nu)\sigma(\nu) \int d\nu' \frac{d Q_\mathrm{fl}(\nu,\nu')}{d\nu'} \\
& = & \nonumber \frac{\Upsilon}{4\pi} \int d\nu \Phi_\mathrm{exc}(\nu)\sigma(\nu) Q_\mathrm{fl}(\nu) \\
& = & \nonumber \frac{\Upsilon}{4\pi} \overline{\sigma}_\mathrm{fl}
\end{eqnarray}
where the right hand side is the integrated BL photon flux due to a given 
molecule, observed along a given line of sight and
$\displaystyle \overline{\sigma}_\mathrm{fl}$ is defined by the equation above.

The BL spectrum in the RR has been reported by \citet{vij04} 
and \citet{vij05}. If a fraction $\eta$ of the total, integrated BL photon flux 
observed is due to a given molecule, we can solve Eq.~(\ref{bl_brightness_3}) 
with respect to $\displaystyle \frac{\Upsilon}{4\pi}$, to yield
\begin{equation}
\frac{\Upsilon}{4\pi} = \frac{\eta}{\overline{\sigma}_\mathrm{fl}} 
\int d\nu \frac{dF_\mathrm{bl}(\nu)}{d\Omega}.
\end{equation}
This result can be plugged back in Eqs.~(\ref{IR_brightness_2}) and
(\ref{ph_brightness_2}), from which we obtain respectively
\begin{equation} \label{main_equation}
\frac{dF_\mathrm{IR}(\nu')}{d\Omega} = \frac{d \overline{\sigma}_\mathrm{IR}(\nu')}{d\nu'}
\frac{{\eta}}{{\overline{\sigma}}_\mathrm{{fl}}} 
\int d\nu \frac{dF_\mathrm{{bl}}{(\nu)}}{d\Omega}
\end{equation}
\begin{equation} \label{main_ph_equation}
{\frac{dF_\mathrm{{ph}}{(\nu')}}{d\Omega} = \frac{d \overline{\sigma}_\mathrm{{ph}}{(\nu')}}{{d\nu'}}
\frac{{\eta}}{\overline{\sigma}_\mathrm{fl}} 
\int d\nu \frac{dF_\mathrm{bl}(\nu)}{d\Omega}.}
\end{equation}
Integrating Eq.~(\ref{main_ph_equation}) and proceeding as in 
Eq.~(\ref{bl_brightness_3}) we obtain:
\begin{eqnarray} \label{ph_brightness_3}
{\int d\nu' \frac{dF_\mathrm{ph}(\nu')}{d\Omega} }& {=} & {
\int d\nu' \frac{d \overline{\sigma}_\mathrm{ph}(\nu')}{d\nu'}
\frac{\eta}{\overline{\sigma}_\mathrm{fl}} 
\int d\nu \frac{dF_\mathrm{bl}(\nu)}{d\Omega} }\\
\nonumber & {=} & {\eta \frac{\overline{\sigma}_\mathrm{ph}}{\overline{\sigma}_\mathrm{fl}}
\int d\nu \frac{dF_\mathrm{bl}(\nu)}{d\Omega}.}
\end{eqnarray}
The ratio $\displaystyle \frac{\overline{\sigma}_\mathrm{ph}}
{\overline{\sigma}_\mathrm{fl}}$, for a given molecule in a given RF, is 
simply the ratio of the branching ratios for phosphorescence and 
fluorescence, as listed in Table~\ref{branching}. The quantities 
$\displaystyle \frac{d \overline{\sigma}_\mathrm{IR}^\mathrm{lobe}(\nu')}{d\nu'}$ 
and $\displaystyle \frac{d \overline{\sigma}_\mathrm{IR}^\mathrm{halo}(\nu')}{d\nu'}$ 
are part of the results of our Monte\textendash Carlo model. The integrated BL 
photon flux along the lines of sight at offsets of 2.5''S~2.6''E 
and 2.5''S~7.8''E from the central source of the RR nebula were obtained
from \citet{vij05} and \citet{vij04}, assuming an average energy of the BL
photons of $\sim$3.1~eV. They are respectively 
${\sim9.2\,\,10^{9}}$~photons~s$^{-1}$~cm$^{-2}$~sr$^{-1}$ at 2.5''S~2.6''E and 
${\sim2.5\,\,10^{9}}$~photons~s$^{-1}$~cm$^{-2}$~sr$^{-1}$ at 2.5''S~7.8''E. 

The line of sight at the offset 2.5''S~7.8''E is completely in the 
halo of the RR nebula. Therefore in this case, with the exception of $\eta$, 
all quantities on the right hand side of Eqs.~(\ref{main_equation}) and
(\ref{main_ph_equation}) are known or can be calculated. We can thus 
quantitatively estimate the expected IR emission and phosphorescence 
spectrum of anthracene, phenanthrene and pyrene apart from a scaling factor 
$\eta$. 

The line of sight at the offset 2.5''S~2.6''E traverses both the southern 
lobe and the halo of the RR nebula. Given the geometry of the source 
\citep{men02}, the part in the halo of this latter line of sight is very 
similar to the line of sight at the offset 2.5''S~7.8''E; we hence assume 
their contributions to both the BL 
(${\sim2.5\,\,10^{9}}$~photons~s$^{-1}$~cm$^{-2}$~sr$^{-1}$) and the IR emission to 
be approximately the same. The remaining BL at 2.5''S~2.6''E
($\sim6.7\,\,10^{9}$~photons~s$^{-1}$~cm$^{-2}$~sr$^{-1}$), after subtracting such 
estimated contribution from the halo, is due to the portion of the line of 
sight through the lobe. Therefore for this part we can again use 
Eq.~(\ref{main_equation}), with the RF in the lobes. The total estimated IR 
emission spectra at 2.5''S~2.6''E can therefore be obtained from the sum of 
the estimated spectra from the portions of the line of sight respectively 
through the lobe and through the halo.

\section{Results}

\subsection{IR emission}

Tables~\ref{anthracene_4-31G}, \ref{anthracene_cc-pvdz}, 
\ref{phenanthrene_4-31G} and \ref{pyrene_4-31G}  list the most intense mid\textendash{} 
and far\textendash IR emission bands expected, along with 
their calculated fluxes, respectively for anthracene (using two different 
vibrational analyses), phenanthrene and pyrene, 
assuming each of them, in turn, to be the sole carrier (i.~e. $\eta\simeq1$) 
of the observed BL at positions 2.5''S~2.6''E and 2.5''S~7.8''E from 
the central source. These fluxes can be simply scaled by the appropriate 
factor if the given molecule is assumed to only contribute a fraction of 
the luminescence. Since the IR fluxes scaled in this way still do 
slightly depend on the assumed photon absorption rate (see the discussion in 
the previous section), we ran independent Monte\textendash Carlo simulations assuming a 
range of different photon absorption rates. In the halo the photon absorption 
rates turn out to be always small enough to allow almost all vibrational 
relaxation cascades to finish. This means that the estimated IR emission from 
the halo is independent of the three\textendash dimensional distribution of the molecules
along the line of sight. 
The situation is slightly different for the lobes. For the closest
position to the central source along the line of sight at 2.5''S~2.6''E we
estimated an average time between absorptions $\tau_\mathrm{abs}\sim{70}$~s 
for pyrene and $\tau_\mathrm{abs}\sim{60}$~s for the other two molecules. We computed 
the estimated IR emission fluxes for the portion in the lobe of the line of 
sight at 2.5''S~2.6''E assuming the above absorption rates and those given by 
an RF dilution by a factor of $\sim6$, corresponding to molecules distributed 
farther away from the nebular axis, near the edges of the lobe. It turns out 
that only the very weakest IR\textendash active bands of each molecule are significantly 
affected, being somewhat suppressed (up to a factor $\sim$1.6 in the worst 
case) when using the higher photon absorption rates.
For all the other IR\textendash active bands, the resulting variation in the estimated 
fluxes is smaller than the uncertainty of the calculated vibrational 
transition intensities we used. 
To evaluate the impact of the latter uncertainty, we performed two 
sets of simulations for anthracene, using the DFT vibrational analyses 
respectively at the B3LYP/4\textendash31G level of theory \citep{lan96} for one and 
at the B3LYP/cc-pvdz \citep{mar96} for the other. 
The latter is very nearly the best currently feasible for 
these molecules, while the former is well known to provide a very good
compromise between accuracy and computational efficiency \citep{lan96,bau97}. 
The overall agreement between the two calculations is very good: while the 
two calculations distribute intensities in a slightly different way among 
nearby vibrational modes, their total is almost the same (see e.~g. the 
in\textendash plane C\textendash H stretches between 3.25 and 3.29~$\mu$m). The differences on 
single band intensities and positions are comparable to the 
relative inaccuracy of each of the two upon comparison with available 
experimental data \citep{hud98}. 

The resulting IR emission fluxes are shown 
respectively in Tables~\ref{anthracene_4-31G} and \ref{anthracene_cc-pvdz}. 
Upon comparison, they show the accuracy of the vibrational analyses to dominate
the uncertainty in the estimated IR emission fluxes.

\begin{table}
\caption{Absolute integrated IR emission fluxes expected for the most intense 
IR\textendash active bands of anthracene, calculated using the vibrational analysis 
performed at the B3LYP/4\textendash31G level of theory. They were estimated for two
offsets from the central source, with two different RF dilution factors 
for the position on the lobe (see text for details).}
\label{anthracene_4-31G}
\begin{center}
\begin{tabular}{ccccc}
\hline \hline \noalign{\smallskip}
Band pos. ($\mu$m) & \multicolumn{3}{c}{Integrated flux/$\eta$ 
($\mathrm{W} \mathrm{sr}^{-1} \mathrm{cm}^{-2}$)} \\
\noalign{\smallskip} & \multicolumn{2}{c}{2.5''S 2.6''E} & 2.5''S 7.8''E \\
& \multicolumn{2}{c}{73\% lobe + 27\% halo} & halo \\
& high abs. rate & low abs. rate & \\
\noalign{\smallskip} \hline \noalign{\smallskip}
3.25 & $1.1\,\,10^{-8}$ & $1.0\,\,10^{-8}$ & $2.0\,\,10^{-9}$ \\
3.26 & $6.9\,\,10^{-9}$ & $6.7\,\,10^{-9}$ & $1.3\,\,10^{-9}$ \\
3.28 & $2.4\,\,10^{-11}$ & $2.2\,\,10^{-11}$ & $4.3\,\,10^{-12}$ \\
3.28 & $1.8\,\,10^{-9}$ & $1.8\,\,10^{-9}$ & $3.5\,\,10^{-10}$ \\
3.29 & $1.0\,\,10^{-9}$ & $1.0\,\,10^{-9}$ & $1.9\,\,10^{-10}$ \\
6.16 & $1.5\,\,10^{-9}$ & $1.5\,\,10^{-9}$ & $3.0\,\,10^{-10}$ \\
6.51 & $4.2\,\,10^{-10}$ & $4.2\,\,10^{-10}$ & $8.9\,\,10^{-11}$ \\
6.86 & $8.3\,\,10^{-10}$ & $8.5\,\,10^{-10}$ & $1.8\,\,10^{-10}$ \\
6.86 & $3.9\,\,10^{-10}$ & $3.9\,\,10^{-10}$ & $8.2\,\,10^{-11}$ \\
7.22 & $3.0\,\,10^{-11}$ & $2.9\,\,10^{-11}$ & $6.0\,\,10^{-12}$ \\
7.43 & $5.4\,\,10^{-10}$ & $5.3\,\,10^{-10}$ & $1.1\,\,10^{-10}$ \\
7.60 & $8.7\,\,10^{-10}$ & $8.5\,\,10^{-10}$ & $1.8\,\,10^{-10}$ \\
7.85 & $1.1\,\,10^{-9}$ & $1.1\,\,10^{-9}$ & $2.4\,\,10^{-10}$ \\
8.55 & $1.5\,\,10^{-10}$ & $1.6\,\,10^{-10}$ & $3.3\,\,10^{-11}$ \\
8.62 & $4.0\,\,10^{-10}$ & $4.0\,\,10^{-10}$ & $8.5\,\,10^{-11}$ \\
8.67 & $1.0\,\,10^{-9}$ & $1.0\,\,10^{-9}$ & $2.1\,\,10^{-10}$ \\
9.95 & $4.8\,\,10^{-10}$ & $4.8\,\,10^{-10}$ & $1.1\,\,10^{-10}$ \\
10.41 & $8.6\,\,10^{-10}$ & $8.7\,\,10^{-10}$ & $1.9\,\,10^{-10}$ \\
11.00 & $1.9\,\,10^{-10}$ & $1.9\,\,10^{-10}$ & $4.0\,\,10^{-11}$ \\
11.32 & $6.3\,\,10^{-9}$ & $6.3\,\,10^{-9}$ & $1.4\,\,10^{-9}$ \\
12.56 & $2.9\,\,10^{-12}$ & $3.4\,\,10^{-12}$ & $2.8\,\,10^{-12}$ \\
13.71 & $6.2\,\,10^{-9}$ & $6.3\,\,10^{-9}$ & $1.4\,\,10^{-9}$ \\
15.33 & $1.1\,\,10^{-10}$ & $1.1\,\,10^{-10}$ & $2.6\,\,10^{-11}$ \\
16.34 & $5.0\,\,10^{-10}$ & $5.1\,\,10^{-10}$ & $1.1\,\,10^{-10}$ \\
21.26 & $8.9\,\,10^{-10}$ & $9.0\,\,10^{-10}$ & $2.0\,\,10^{-10}$ \\
26.35 & $2.1\,\,10^{-11}$ & $2.9\,\,10^{-11}$ & $1.7\,\,10^{-11}$ \\
43.68 & $9.1\,\,10^{-11}$ & $1.3\,\,10^{-10}$ & $3.3\,\,10^{-11}$ \\
110.23 & $5.0\,\,10^{-11}$ & $7.5\,\,10^{-11}$ & $4.0\,\,10^{-11}$ \\
\noalign{\smallskip} \hline \hline
\end{tabular}
\end{center}
\end{table}

\begin{table}
\caption{Absolute integrated IR emission fluxes expected for the most intense 
IR\textendash active bands of anthracene, calculated using the vibrational analysis 
performed at the B3LYP/cc\textendash pvdz level of theory \citep{mar96}. They were 
estimated for two offsets from the central source, with two different RF 
dilution factors for the position on the lobe (see text for details).}
\label{anthracene_cc-pvdz}
\begin{center}
\begin{tabular}{ccccc}
\hline \hline \noalign{\smallskip}
Band pos. ($\mu$m) & \multicolumn{3}{c}{Integrated flux/$\eta$ 
($\mathrm{W} \mathrm{sr}^{-1} \mathrm{cm}^{-2}$)} \\
\noalign{\smallskip} & \multicolumn{2}{c}{2.5''S 2.6''E} & 2.5''S 7.8''E \\
& \multicolumn{2}{c}{73\% lobe + 27\% halo} & halo \\
& high abs. rate & low abs. rate & \\
\noalign{\smallskip} \hline \noalign{\smallskip}
3.26 & $8.1\,\,10^{-9}$ & $8.1\,\,10^{-9}$ & $1.5\,\,10^{-9}$ \\
3.27 & $9.0\,\,10^{-9}$ & $9.1\,\,10^{-9}$ & $1.7\,\,10^{-9}$ \\
3.29 & $1.7\,\,10^{-9}$ & $1.7\,\,10^{-9}$ & $3.2\,\,10^{-10}$ \\
3.29 & $1.4\,\,10^{-9}$ & $1.3\,\,10^{-9}$ & $2.8\,\,10^{-10}$ \\
6.14 & $1.0\,\,10^{-9}$ & $1.1\,\,10^{-9}$ & $2.2\,\,10^{-10}$ \\
6.50 & $8.8\,\,10^{-10}$ & $9.1\,\,10^{-10}$ & $1.8\,\,10^{-10}$ \\
6.94 & $1.7\,\,10^{-10}$ & $1.8\,\,10^{-10}$ & $3.7\,\,10^{-11}$ \\
6.98 & $1.7\,\,10^{-10}$ & $1.8\,\,10^{-10}$ & $3.7\,\,10^{-11}$ \\
7.13 & $3.4\,\,10^{-10}$ & $3.4\,\,10^{-10}$ & $7.2\,\,10^{-11}$ \\
7.34 & $5.0\,\,10^{-10}$ & $5.2\,\,10^{-10}$ & $1.1\,\,10^{-10}$ \\
7.71 & $9.9\,\,10^{-10}$ & $1.0\,\,10^{-9}$ & $2.1\,\,10^{-10}$ \\
7.94 & $9.9\,\,10^{-10}$ & $1.0\,\,10^{-9}$ & $2.1\,\,10^{-10}$ \\
8.61 & $3.1\,\,10^{-10}$ & $3.3\,\,10^{-10}$ & $6.7\,\,10^{-11}$ \\
8.73 & $1.1\,\,10^{-9}$ & $1.1\,\,10^{-9}$ & $2.3\,\,10^{-10}$ \\
8.93 & $3.1\,\,10^{-10}$ & $3.1\,\,10^{-10}$ & $6.8\,\,10^{-11}$ \\
10.01 & $8.6\,\,10^{-10}$ & $9.0\,\,10^{-10}$ & $1.8\,\,10^{-10}$ \\
10.50 & $5.6\,\,10^{-10}$ & $5.8\,\,10^{-10}$ & $1.2\,\,10^{-10}$ \\
11.15 & $2.6\,\,10^{-10}$ & $2.8\,\,10^{-10}$ & $5.9\,\,10^{-11}$ \\
11.35 & $4.9\,\,10^{-9}$ & $5.2\,\,10^{-9}$ & $1.1\,\,10^{-9}$ \\
13.76 & $6.1\,\,10^{-9}$ & $6.4\,\,10^{-9}$ & $1.3\,\,10^{-9}$ \\
15.43 & $1.1\,\,10^{-10}$ & $1.3\,\,10^{-10}$ & $2.6\,\,10^{-11}$ \\
16.56 & $7.4\,\,10^{-10}$ & $7.8\,\,10^{-10}$ & $1.6\,\,10^{-10}$ \\
21.28 & $1.1\,\,10^{-9}$ & $1.2\,\,10^{-9}$ & $2.6\,\,10^{-10}$ \\
43.29 & $6.9\,\,10^{-11}$ & $1.1\,\,10^{-10}$ & $3.0\,\,10^{-11}$ \\
111.11 & $4.9\,\,10^{-11}$ & $7.4\,\,10^{-11}$ & $3.9\,\,10^{-11}$ \\
\noalign{\smallskip} \hline \hline
\end{tabular}
\end{center}
\end{table}

\begin{table}
\caption{Absolute integrated IR emission fluxes expected for the most intense 
IR\textendash active bands of phenanthrene, calculated using the vibrational analysis 
performed at the B3LYP/4\textendash31G level of theory. They were estimated for two
offsets from the central source, with two different RF dilution factors 
for the position on the lobe (see text for details).}
\label{phenanthrene_4-31G}
\begin{center}
\begin{tabular}{ccccc}
\hline \hline \noalign{\smallskip}
Band pos. ($\mu$m) & \multicolumn{3}{c}{Integrated flux/$\eta$ 
($\mathrm{W} \mathrm{sr}^{-1} \mathrm{cm}^{-2}$)} \\
\noalign{\smallskip} & \multicolumn{2}{c}{2.5''S 2.6''E} & 2.5''S 7.8''E \\
& \multicolumn{2}{c}{73\% lobe + 27\% halo} & halo \\
& high abs. rate & low abs. rate & \\
\noalign{\smallskip} \hline \noalign{\smallskip}
3.23 & $1.3\,\,10^{-8}$ & $1.3\,\,10^{-8}$ & $1.9\,\,10^{-9}$ \\
3.24 & $1.5\,\,10^{-8}$ & $1.5\,\,10^{-8}$ & $2.3\,\,10^{-9}$ \\
3.25 & $1.2\,\,10^{-9}$ & $1.2\,\,10^{-9}$ & $1.6\,\,10^{-10}$ \\
3.26 & $2.1\,\,10^{-8}$ & $2.0\,\,10^{-8}$ & $3.1\,\,10^{-9}$ \\
3.26 & $1.9\,\,10^{-8}$ & $1.8\,\,10^{-8}$ & $2.8\,\,10^{-9}$ \\
3.27 & $8.1\,\,10^{-9}$ & $7.9\,\,10^{-9}$ & $1.2\,\,10^{-9}$ \\
3.27 & $2.8\,\,10^{-11}$ & $3.0\,\,10^{-11}$ & $3.0\,\,10^{-12}$ \\
3.28 & $1.1\,\,10^{-9}$ & $1.1\,\,10^{-9}$ & $1.7\,\,10^{-10}$ \\
3.28 & $2.1\,\,10^{-9}$ & $2.1\,\,10^{-9}$ & $3.4\,\,10^{-10}$ \\
3.29 & $2.2\,\,10^{-10}$ & $2.3\,\,10^{-10}$ & $3.8\,\,10^{-11}$ \\
6.21 & $4.1\,\,10^{-10}$ & $4.2\,\,10^{-10}$ & $6.7\,\,10^{-11}$ \\
6.23 & $2.9\,\,10^{-10}$ & $2.8\,\,10^{-10}$ & $4.7\,\,10^{-11}$ \\
6.27 & $2.3\,\,10^{-9}$ & $2.2\,\,10^{-9}$ & $3.7\,\,10^{-10}$ \\
6.43 & $6.3\,\,10^{-12}$ & $2.9\,\,10^{-12}$ & $7.7\,\,10^{-13}$ \\
6.57 & $9.1\,\,10^{-10}$ & $8.9\,\,10^{-10}$ & $1.4\,\,10^{-10}$ \\
6.68 & $3.6\,\,10^{-9}$ & $3.6\,\,10^{-9}$ & $6.1\,\,10^{-10}$ \\
6.84 & $7.2\,\,10^{-9}$ & $7.1\,\,10^{-9}$ & $1.2\,\,10^{-9}$ \\
6.93 & $1.8\,\,10^{-9}$ & $1.8\,\,10^{-9}$ & $3.1\,\,10^{-10}$ \\
7.04 & $3.3\,\,10^{-10}$ & $3.1\,\,10^{-10}$ & $5.4\,\,10^{-11}$ \\
7.06 & $4.7\,\,10^{-10}$ & $4.7\,\,10^{-10}$ & $7.9\,\,10^{-11}$ \\
7.45 & $1.1\,\,10^{-9}$ & $1.1\,\,10^{-9}$ & $1.9\,\,10^{-10}$ \\
7.49 & $2.3\,\,10^{-11}$ & $2.4\,\,10^{-11}$ & $4.9\,\,10^{-12}$ \\
7.70 & $7.9\,\,10^{-10}$ & $7.9\,\,10^{-10}$ & $1.3\,\,10^{-10}$ \\
7.76 & $2.6\,\,10^{-11}$ & $3.0\,\,10^{-11}$ & $5.3\,\,10^{-12}$ \\
8.00 & $4.6\,\,10^{-9}$ & $4.6\,\,10^{-9}$ & $7.9\,\,10^{-10}$ \\
8.16 & $4.4\,\,10^{-10}$ & $4.2\,\,10^{-10}$ & $7.3\,\,10^{-11}$ \\
8.31 & $9.6\,\,10^{-10}$ & $9.8\,\,10^{-10}$ & $1.6\,\,10^{-10}$ \\
8.46 & $3.1\,\,10^{-10}$ & $3.3\,\,10^{-10}$ & $5.5\,\,10^{-11}$ \\
8.52 & $1.4\,\,10^{-11}$ & $1.4\,\,10^{-11}$ & $2.9\,\,10^{-12}$ \\
8.60 & $9.6\,\,10^{-11}$ & $1.1\,\,10^{-10}$ & $2.0\,\,10^{-11}$ \\
8.71 & $5.8\,\,10^{-10}$ & $6.1\,\,10^{-10}$ & $1.0\,\,10^{-10}$ \\
9.15 & $3.9\,\,10^{-10}$ & $4.0\,\,10^{-10}$ & $6.8\,\,10^{-11}$ \\
9.63 & $1.6\,\,10^{-9}$ & $1.6\,\,10^{-9}$ & $2.9\,\,10^{-10}$ \\
9.66 & $2.9\,\,10^{-10}$ & $2.9\,\,10^{-10}$ & $5.1\,\,10^{-11}$ \\
10.01 & $6.1\,\,10^{-10}$ & $6.2\,\,10^{-10}$ & $1.1\,\,10^{-10}$ \\
10.12 & $2.4\,\,10^{-11}$ & $3.4\,\,10^{-11}$ & $6.5\,\,10^{-12}$ \\
10.53 & $1.9\,\,10^{-9}$ & $1.8\,\,10^{-9}$ & $3.4\,\,10^{-10}$ \\
11.48 & $4.4\,\,10^{-9}$ & $4.5\,\,10^{-9}$ & $7.9\,\,10^{-10}$ \\
11.49 & $6.4\,\,10^{-10}$ & $6.2\,\,10^{-10}$ & $1.2\,\,10^{-10}$ \\
12.05 & $4.7\,\,10^{-11}$ & $6.5\,\,10^{-11}$ & $1.6\,\,10^{-11}$ \\
12.24 & $2.2\,\,10^{-8}$ & $2.3\,\,10^{-8}$ & $4.0\,\,10^{-9}$ \\
13.58 & $2.6\,\,10^{-8}$ & $2.6\,\,10^{-8}$ & $4.8\,\,10^{-9}$ \\
13.95 & $8.6\,\,10^{-10}$ & $9.1\,\,10^{-10}$ & $1.7\,\,10^{-10}$ \\
13.98 & $5.0\,\,10^{-10}$ & $5.1\,\,10^{-10}$ & $9.7\,\,10^{-11}$ \\
14.12 & $5.2\,\,10^{-11}$ & $8.4\,\,10^{-11}$ & $2.5\,\,10^{-11}$ \\
15.94 & $1.4\,\,10^{-9}$ & $1.5\,\,10^{-9}$ & $2.8\,\,10^{-10}$ \\
18.19 & $1.9\,\,10^{-10}$ & $2.5\,\,10^{-10}$ & $6.0\,\,10^{-11}$ \\
19.99 & $3.2\,\,10^{-10}$ & $3.8\,\,10^{-10}$ & $8.4\,\,10^{-11}$ \\
20.07 & $1.1\,\,10^{-9}$ & $1.1\,\,10^{-9}$ & $2.2\,\,10^{-10}$ \\
22.75 & $5.4\,\,10^{-10}$ & $6.1\,\,10^{-10}$ & $1.3\,\,10^{-10}$ \\
23.25 & $1.2\,\,10^{-9}$ & $1.3\,\,10^{-9}$ & $2.6\,\,10^{-10}$ \\
24.72 & $2.1\,\,10^{-10}$ & $3.2\,\,10^{-10}$ & $8.3\,\,10^{-11}$ \\
41.08 & $2.1\,\,10^{-10}$ & $3.5\,\,10^{-10}$ & $1.3\,\,10^{-10}$ \\
44.26 & $5.5\,\,10^{-10}$ & $7.1\,\,10^{-10}$ & $1.7\,\,10^{-10}$ \\
100.13 & $2.3\,\,10^{-10}$ & $3.3\,\,10^{-10}$ & $1.9\,\,10^{-10}$ \\
\noalign{\smallskip} \hline \hline
\end{tabular}
\end{center}
\end{table}

\begin{table}
\caption{Absolute integrated IR emission fluxes expected for the most intense 
IR\textendash active bands of pyrene, calculated using the vibrational analysis 
performed at the B3LYP/4\textendash31G level of theory. They were estimated for two
offsets from the central source, with two different RF dilution factors 
for the position on the lobe (see text for details).}
\label{pyrene_4-31G}
\begin{center}
\begin{tabular}{ccccc}
\hline \hline \noalign{\smallskip}
Band pos. ($\mu$m) & \multicolumn{3}{c}{Integrated flux/$\eta$ 
($\mathrm{W} \mathrm{sr}^{-1} \mathrm{cm}^{-2}$)} \\
\noalign{\smallskip} & \multicolumn{2}{c}{2.5''S 2.6''E} & 2.5''S 7.8''E \\
& \multicolumn{2}{c}{73\% lobe + 27\% halo} & halo \\
& high abs. rate & low abs. rate & \\
\noalign{\smallskip} \hline \noalign{\smallskip}
3.25 & $2.0\,\,10^{-8}$ & $2.0\,\,10^{-8}$ & $3.0\,\,10^{-9}$ \\
3.26 & $2.0\,\,10^{-8}$ & $1.9\,\,10^{-8}$ & $2.9\,\,10^{-9}$ \\
3.27 & $5.4\,\,10^{-9}$ & $5.3\,\,10^{-9}$ & $8.0\,\,10^{-10}$ \\
3.28 & $8.3\,\,10^{-12}$ & $8.8\,\,10^{-12}$ & $1.3\,\,10^{-12}$ \\
3.29 & $7.9\,\,10^{-10}$ & $7.7\,\,10^{-10}$ & $1.1\,\,10^{-10}$ \\
6.26 & $1.6\,\,10^{-9}$ & $1.6\,\,10^{-9}$ & $2.6\,\,10^{-10}$ \\
6.31 & $3.7\,\,10^{-9}$ & $3.6\,\,10^{-9}$ & $6.0\,\,10^{-10}$ \\
6.78 & $1.1\,\,10^{-9}$ & $1.1\,\,10^{-9}$ & $1.9\,\,10^{-10}$ \\
6.92 & $1.5\,\,10^{-10}$ & $1.5\,\,10^{-10}$ & $2.3\,\,10^{-11}$ \\
7.01 & $3.2\,\,10^{-10}$ & $3.4\,\,10^{-10}$ & $5.6\,\,10^{-11}$ \\
7.01 & $3.3\,\,10^{-9}$ & $3.3\,\,10^{-9}$ & $5.6\,\,10^{-10}$ \\
7.61 & $1.8\,\,10^{-9}$ & $1.8\,\,10^{-9}$ & $3.0\,\,10^{-10}$ \\
7.98 & $9.8\,\,10^{-10}$ & $1.0\,\,10^{-9}$ & $1.7\,\,10^{-10}$ \\
8.29 & $3.0\,\,10^{-12}$ & $4.5\,\,10^{-12}$ & $1.1\,\,10^{-12}$ \\
8.42 & $2.8\,\,10^{-9}$ & $2.8\,\,10^{-9}$ & $4.7\,\,10^{-10}$ \\
8.62 & $4.3\,\,10^{-10}$ & $4.2\,\,10^{-10}$ & $7.4\,\,10^{-11}$ \\
9.16 & $1.2\,\,10^{-9}$ & $1.3\,\,10^{-9}$ & $2.2\,\,10^{-10}$ \\
10.04 & $1.4\,\,10^{-10}$ & $1.5\,\,10^{-10}$ & $2.6\,\,10^{-11}$ \\
10.25 & $6.5\,\,10^{-10}$ & $6.5\,\,10^{-10}$ & $1.1\,\,10^{-10}$ \\
10.47 & $5.2\,\,10^{-11}$ & $5.5\,\,10^{-11}$ & $1.0\,\,10^{-11}$ \\
11.79 & $2.7\,\,10^{-8}$ & $2.7\,\,10^{-8}$ & $4.8\,\,10^{-9}$ \\
12.20 & $6.8\,\,10^{-10}$ & $7.0\,\,10^{-10}$ & $1.3\,\,10^{-10}$ \\
13.40 & $2.0\,\,10^{-9}$ & $2.0\,\,10^{-9}$ & $3.7\,\,10^{-10}$ \\
14.06 & $6.2\,\,10^{-9}$ & $6.3\,\,10^{-9}$ & $1.1\,\,10^{-9}$ \\
14.43 & $3.9\,\,10^{-11}$ & $5.5\,\,10^{-11}$ & $1.3\,\,10^{-11}$ \\
18.20 & $4.3\,\,10^{-10}$ & $4.4\,\,10^{-10}$ & $8.2\,\,10^{-11}$ \\
20.00 & $4.1\,\,10^{-10}$ & $4.3\,\,10^{-10}$ & $8.1\,\,10^{-11}$ \\
20.38 & $2.7\,\,10^{-10}$ & $2.9\,\,10^{-10}$ & $5.7\,\,10^{-11}$ \\
28.32 & $2.2\,\,10^{-10}$ & $2.7\,\,10^{-10}$ & $5.5\,\,10^{-11}$ \\
47.80 & $4.6\,\,10^{-10}$ & $5.3\,\,10^{-10}$ & $1.0\,\,10^{-10}$ \\
101.60 & $9.2\,\,10^{-11}$ & $1.3\,\,10^{-10}$ & $8.0\,\,10^{-11}$ \\
\noalign{\smallskip} \hline \hline
\end{tabular}
\end{center}
\end{table}

\subsection{Phosphorescence}

Table~\ref{phosphotable} lists the integrated phosphorescence
fluxes expected for phenanthrene and pyrene, again assuming each of 
them, in turn, to be the sole carrier (i.~e. $\eta\simeq1$) of the 
observed BL at positions 2.5''S~2.6''E and 2.5''S~7.8''E from the central 
source. As for the IR fluxes, phosphorescence can be simply scaled by the 
appropriate factor if fluorescence by the given molecule is assumed to 
only contribute a fraction of the BL. Since the branching ratios for 
phosphorescence is extremely small for anthracene in all cases, it is 
omitted in this section. The average phosphorescence photon energy can 
be estimated from the phosphorescence spectra published by \citet{sal04}
to be about 2.4~eV for phenanthrene and 2.0~eV for pyrene respectively. Such 
spectra were taken under experimental conditions specifically chosen to enhance
the phosphorescence yield, in solution, meaning that we can only use them 
as a guide for the spectral profile of the expected phosphorescence of
the same molecules in the RR. Indeed, gas\textendash phase phosphorescence spectra, 
in a collision\textendash free environment, can be expected to arise from highly 
excited vibrational states and therefore have a different width and 
vibronic structure \citep{bre05}. We emphasize that we relied only on the 
gas\textendash phase measurements by \citet{bre05} for the branching ratios, using
Eq.~(\ref{ph_brightness_3}).

\begin{table}
\caption{Absolute integrated phosphorescence fluxes expected for 
phenanthrene and pyrene. They were estimated for two 
offsets from the central source (see text for details).}
\label{phosphotable}
\begin{center}
\begin{tabular}{ccc}
\hline \hline \noalign{\smallskip}
Molecule & \multicolumn{2}{c}{Integrated flux/$\eta$ 
($\mathrm{\textbf{erg}}$~$\mathrm{\textbf{sr}}^{-1}$~$\mathrm{\textbf{cm}}^{-2}$)}
\\
\noalign{\smallskip} & 2.5''S 2.6''E & 2.5''S 7.8''E \\
& 73\% lobe + 27\% halo & halo \\
\noalign{\smallskip} \hline \noalign{\smallskip}
Phenanthrene & 0.438 & 0.130 \\
Pyrene & 0.136 & 0.035 \\
\noalign{\smallskip} \hline \hline
\end{tabular}
\end{center}
\end{table}

\section{Comparison with available ISO data}\label{tps}

\begin{figure}
\begin{center}
\includegraphics{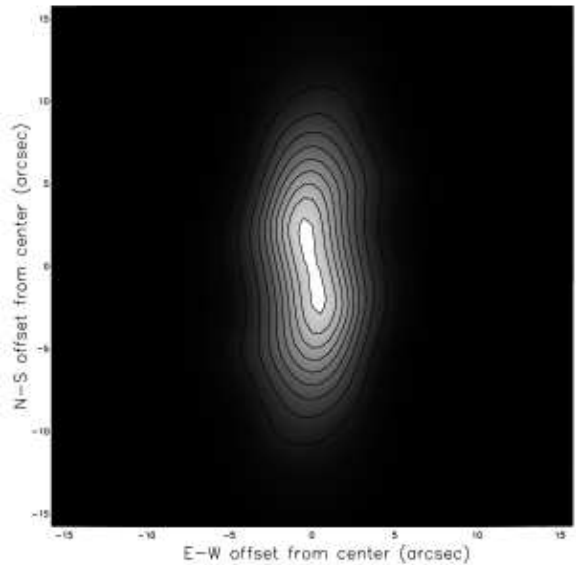}
\end{center}
\caption{Spatial distribution of the surface brightness of the BL, 
interpolated from the values given by \citet{vij05} using a TPS (see text 
for details). East is left, North is up, contour lines are in 5\% intervals.}
\label{tpl_surface}
\end{figure}

\begin{figure}
\begin{center}
\includegraphics{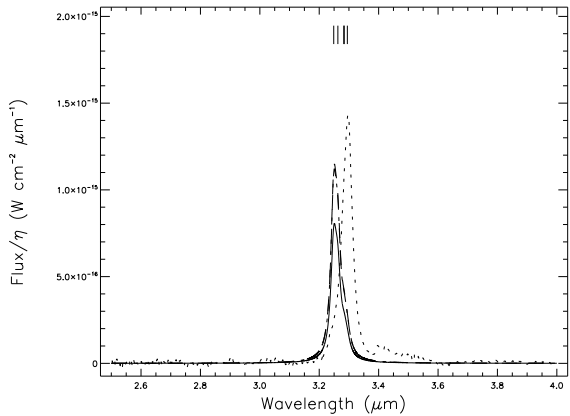}
\end{center}
\caption{Comparison between the estimated IR emission spectrum of anthracene 
(C$_{14}$H$_{10}$) and an ISO spectrum of the RR in the wavelength range 
2.5\textendash4.0~$\mu$m. Calculated spectra, under different assumptions 
(see text for details), are drawn in dashed (calculated from the 
fluxes in second column of Table~\ref{anthracene_4-31G}), dash\textendash dotted 
(third column)} and continuous (fourth column) lines, the 
continuum\textendash subtracted ISO spectrum is shown as a dotted line. The central 
positions of expected anthracene bands are marked by ticks.
\label{anthracene_3.3}
\end{figure}

\begin{figure}
\begin{center}
\includegraphics{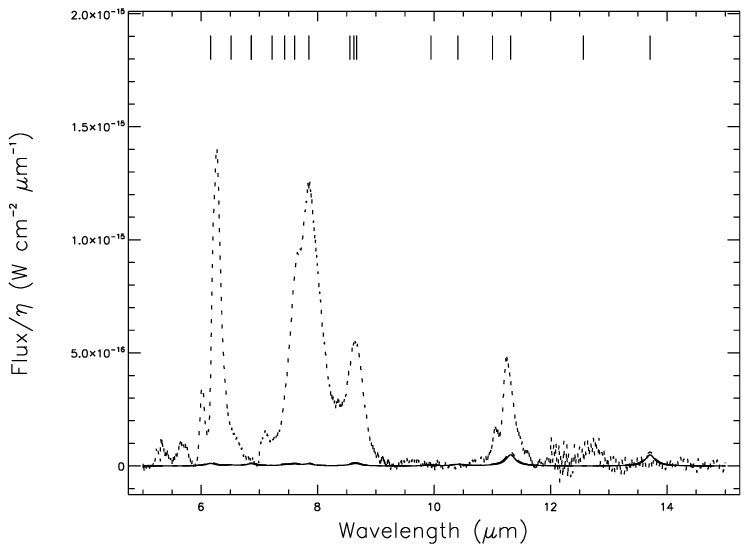}
\end{center}
\caption{Same as Fig.~\ref{anthracene_3.3} in the wavelength range 5\textendash15~$\mu$m.}
\label{anthracene_10}
\end{figure}

\begin{figure}
\begin{center}
\includegraphics{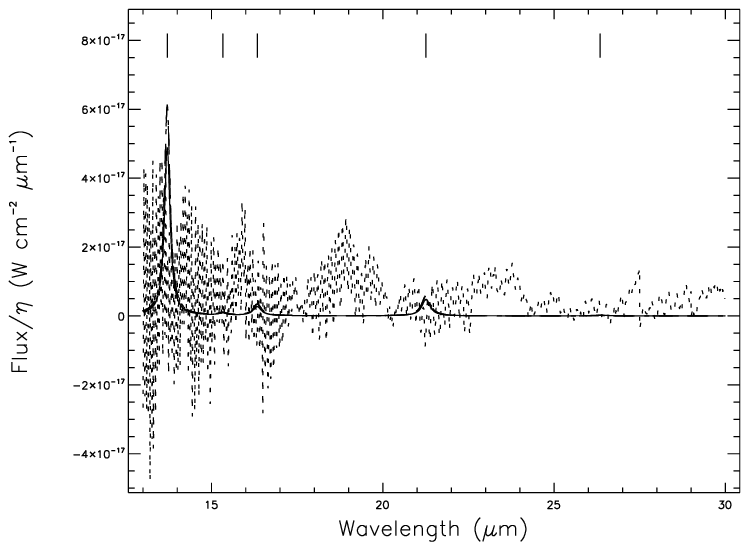}
\end{center}
\caption{Same as Fig.~\ref{anthracene_3.3} in the wavelength range 
13\textendash30~$\mu$m.}
\label{anthracene_22}
\end{figure}

\begin{figure}
\begin{center}
\includegraphics{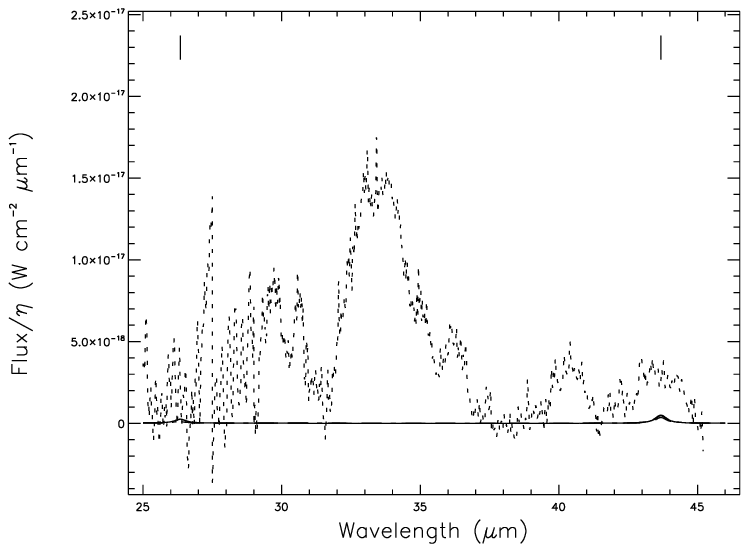}
\end{center}
\caption{Same as Fig.~\ref{anthracene_3.3} in the wavelength range 
25\textendash45~$\mu$m.}
\label{anthracene_45}
\end{figure}

\begin{figure}
\begin{center}
\includegraphics{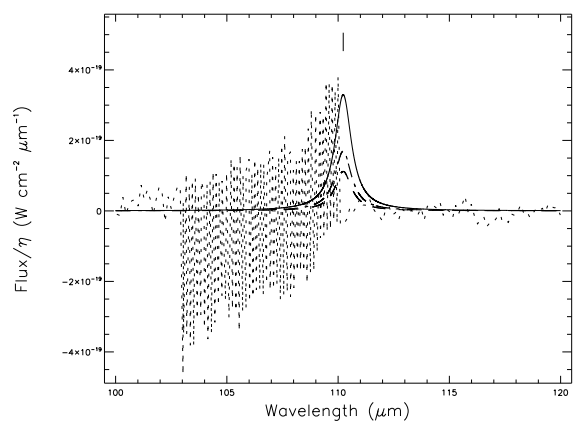}
\end{center}
\caption{Same as Fig.~\ref{anthracene_3.3} in the wavelength range 
100\textendash120~$\mu$m.}
\label{anthracene_110}
\end{figure}

\begin{figure}
\begin{center}
\includegraphics{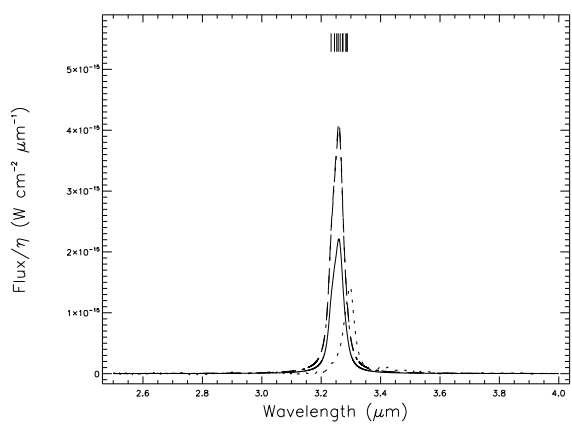}
\end{center}
\caption{Same as Fig.~\ref{anthracene_3.3} for phenanthrene (C$_{14}$H$_{10}$) 
in the wavelength range 2.5\textendash4.0~$\mu$m.}
\label{phenanthrene_3.3}
\end{figure}

\begin{figure}
\begin{center}
\includegraphics{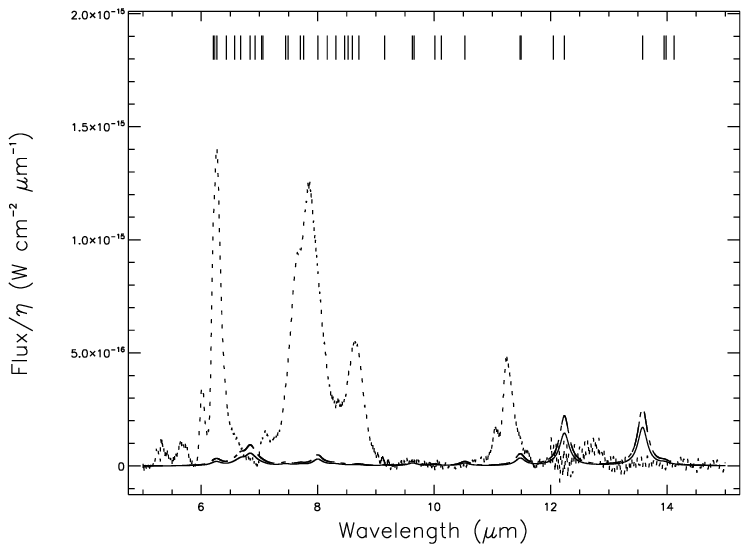}
\end{center}
\caption{Same as Fig.~\ref{phenanthrene_3.3} in the wavelength range 
5\textendash15~$\mu$m.}
\label{phenanthrene_10}
\end{figure}

\begin{figure}
\begin{center}
\includegraphics{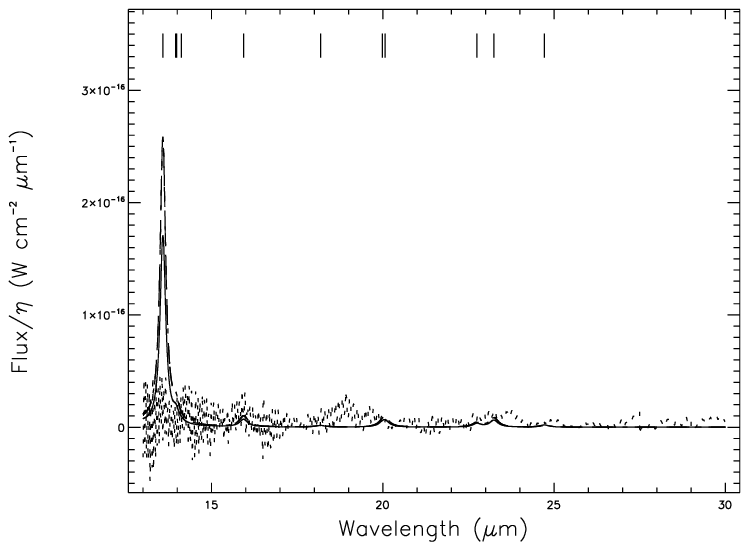}
\end{center}
\caption{Same as Fig.~\ref{phenanthrene_3.3} in the wavelength range 
13\textendash30~$\mu$m.}
\label{phenanthrene_22}
\end{figure}

\begin{figure}
\begin{center}
\includegraphics{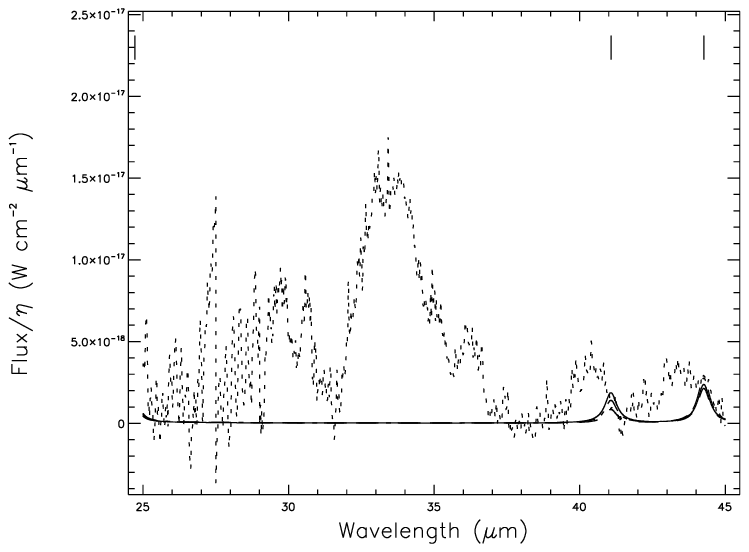}
\end{center}
\caption{Same as Fig.~\ref{phenanthrene_3.3} in the wavelength range 
25\textendash45~$\mu$m.}
\label{phenanthrene_45}
\end{figure}

\begin{figure}
\begin{center}
\includegraphics{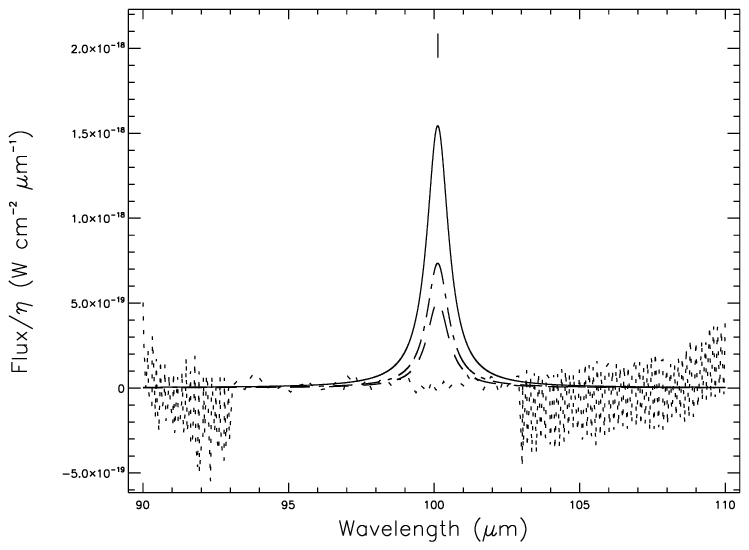}
\end{center}
\caption{Same as Fig.~\ref{phenanthrene_3.3} in the wavelength range 
90\textendash110~$\mu$m.}
\label{phenanthrene_100}
\end{figure}

\begin{figure}
\begin{center}
\includegraphics{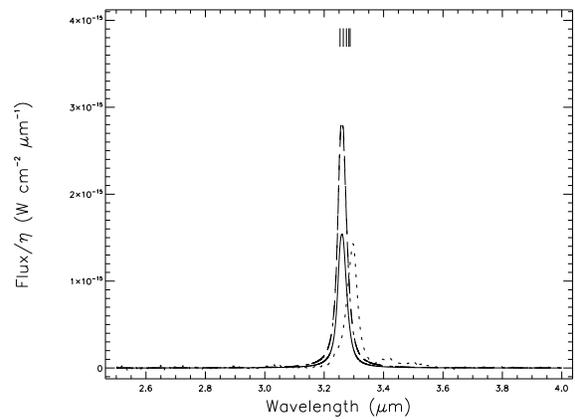}
\end{center}
\caption{Same as Fig.~\ref{anthracene_3.3} for pyrene (C$_{16}$H$_{10}$) in the 
wavelength range 2.5\textendash4.0~$\mu$m.}
\label{pyrene_3.3}
\end{figure}

\begin{figure}
\begin{center}
\includegraphics{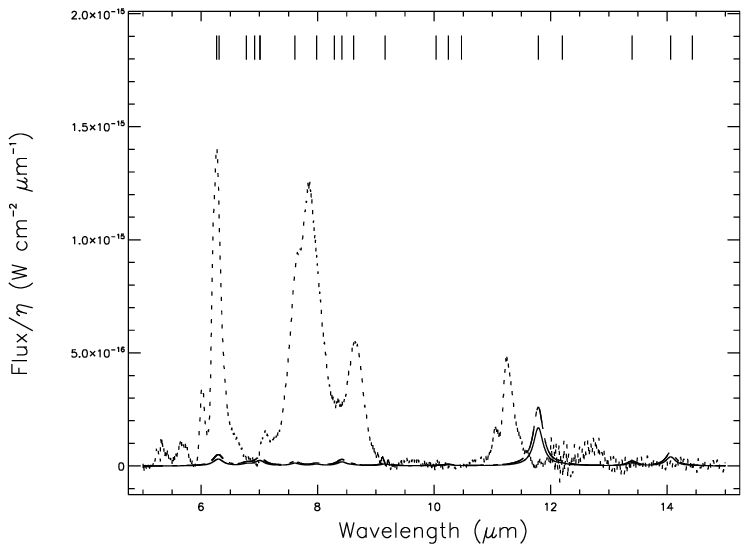}
\end{center}
\caption{Same as Fig.~\ref{pyrene_3.3} in the wavelength range 5\textendash15~$\mu$m.}
\label{pyrene_10}
\end{figure}

\begin{figure}
\begin{center}
\includegraphics{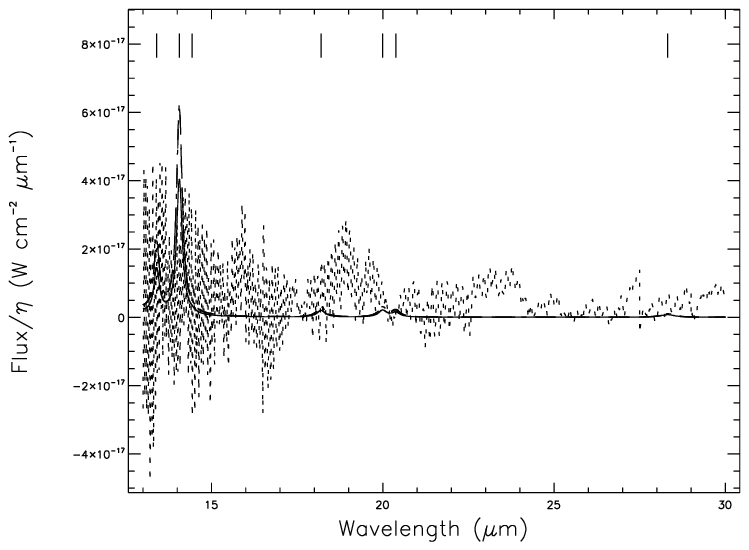}
\end{center}
\caption{Same as Fig.~\ref{pyrene_3.3} in the wavelength range 13\textendash30~$\mu$m.}
\label{pyrene_22}
\end{figure}

\begin{figure}
\begin{center}
\includegraphics{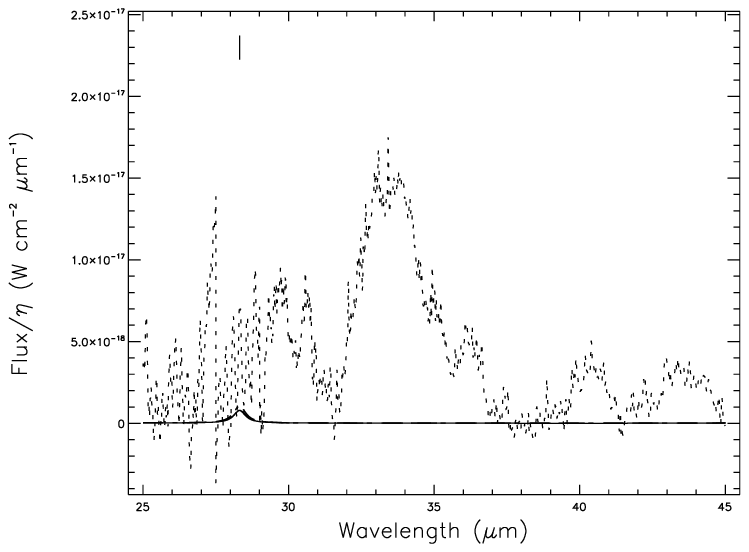}
\end{center}
\caption{Same as Fig.~\ref{pyrene_3.3} in the wavelength range 25\textendash45~$\mu$m.}
\label{pyrene_45}
\end{figure}

\begin{figure}
\begin{center}
\includegraphics{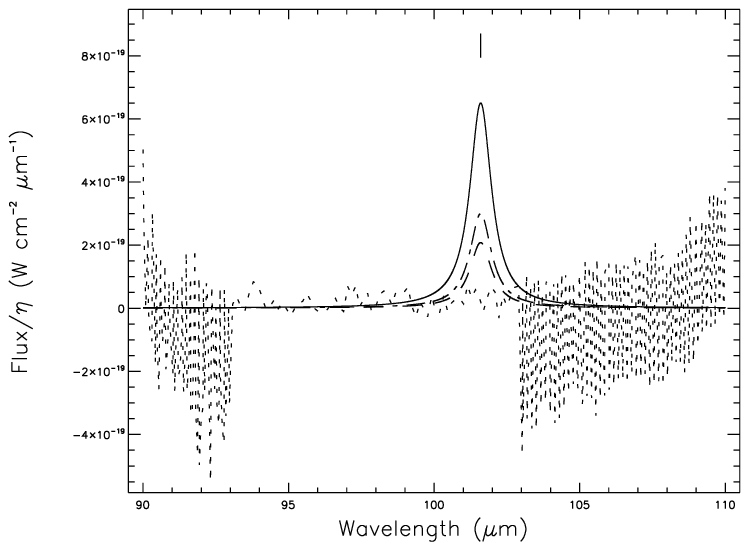}
\end{center}
\caption{Same as Fig.~\ref{pyrene_3.3} in the wavelength range 90\textendash110~$\mu$m.}
\label{pyrene_100}
\end{figure}

The RR nebula has been observed several times with both the Short 
Wavelength Spectrometer (SWS) and the Long Wavelength Spectrometer (LWS)
of ISO. The resulting spectra are available through the online
ISO database. All of these spectra were taken with an entrance slit which
includes by and large the whole RR nebula, e.~g. $33''\times20''$ for the SWS 
and a circular aperture of $84''$ diameter for LWS. Therefore, to compare
our estimated fluxes with ISO observations we would need to calculate 
the predicted IR emission spectrum on a grid of points adequately sampling
the aperture on the sky of the appropriate ISO instrument. This, in turn, 
would require BL measurements on such a grid, which are not available to 
date. To get a (rough) estimate of the BL distribution, we interpolated
the available BL measurements from \citet{vij05} with a thin plate spline
(TPS). A TPS is a smooth function and passes through the original points 
$f(x_i,y_i) = z_i$. It also has the property to become almost linear in the 
independent variables when away from the points used to define it. It 
resembles the more commonly used cubic (or more generally polynomial) 
splines, with the additional qualities of being smooth (and not only 
continuous up to a given derivative order) and naturally multidimensional, 
hence well suited to provide a representation of a 2D quantity as the BL 
surface brightness on the sky. A TPS is defined as
\begin{equation}\label{TPS_def1}
f(x,y) = a_0+a_1x+a_2y+\frac{1}{2}\sum_{i=1}^{n}b_ir_i^2\log(r_i^2), 
\end{equation}
with the constraints
\begin{equation}\label{TPS_def2}
\sum_{i=1}^{n}b_i = \sum_{i=1}^{n}b_ix_i = \sum_{i}^{n}b_iy_i = 0, 
\end{equation}
where $r_i^2 = (x-x_i)^2+(y-y_i)^2$. 
We used the implementation of the TPS provided by the Interactive Data 
Language (IDL) environment. Since \citet{vij05} provided BL measurements on 
two long slits both south of the central source, in the interpolation we 
assumed the BL emission to be symmetric for inversion. In order to obtain a 
sensible asymptotic behaviour of the extrapolated BL, we built the TPS on
the logarithm of the BL surface brightness, instead of the BL itself. 
Hence, in Eqs.~(\ref{TPS_def1}) and (\ref{TPS_def2}) $x$ and $y$ are the
angular offsets from the central source of the RR, and $z$ is the 
logarithm of the corresponding integrated BL.
Figure~\ref{tpl_surface} shows a surface plot of the resulting TPS which, 
despite the small number of data points from which it was obtained, clearly
shows the geometry of the RR nebula. 
We were therefore able to integrate the TPS on the ISO apertures, to yield 
respectively $\sim{1.40\,\,10^{-17}}$~W~cm$^{-2}$ for the $33''\times20''$ 
rectangular aperture of SWS and $\sim{1.45\,\,10^{-17}}$~W~cm$^{-2}$ for the 
circular aperture of diameter $84''$ of LWS. We remark that these integrated
fluxes ought to be considered with caution, given the very sparse spatial 
sampling of the BL. Nonetheless, we used them to scale the calculated
fluxes listed in tables~\ref{anthracene_4-31G}, \ref{phenanthrene_4-31G} and
\ref{pyrene_4-31G}, which were computed for specific positions in the RR
nebula, to these ISO apertures, for comparison with available ISO data.
In particular, to estimate the fluxes expected to be observed by SWS using
the $33''\times20''$ aperture from the fluxes calculated for the RR lobe 
$2.5''$S~$2.6''$E from the central source, we multiplied the latter
by $\sim{1.40\,\,10^{-17}}$~W~cm$^{-2}$ and divided the result by 
$\sim{4.58\,\,10^{-9}}$~W~cm$^{-2}$~sr$^{-1}$ (the BL intensity measured 
by \citet{vij05} at this position in the RR). To ease the comparison,
Figs. \ref{anthracene_3.3} to \ref{pyrene_100} show the resulting
scaled calculated spectra, along with the relevant spectra taken from the 
ISO data archive.

For the SWS we chose the spectrum with the TDT number
70201801, a low\textendash resolution full\textendash grating scan. For the LWS we chose the
spectrum with the TDT number 70901203, a medium resolution spectrum.
In both cases, a cubic spline was fitted to the continuum below the 
emission bands, and subtracted before comparison. No further processing
was performed on the ISO spectra, which therefore are by and large the
automatically reduced spectra as obtained from the ISO data archive tool, 
just continuum\textendash subtracted. In each figure we show the calculated IR emission
bands from a given PAH, respectively scaled from the ``high absorption rate''
and ``low absorption rate'' $2.5''$S~$2.6''$E columns (dashed and dash\textendash dotted
line), and scaled from the $2.5''$S~$7.8''$E column (continuous line). The
continuum\textendash subtracted ISO spectrum is plotted as a dotted line. The positions
of calculated bands are marked by ticks on each plot. Each expected band is
plotted as a lorentzian curve (corresponding to homogeneous band broadening, 
see \citet{pec02}) whose integrated flux is normalised to the expected scaled 
value. The widths of the lorentzians were arbitrarily chosen to match those of 
the observed bands in the same spectral range. In particular, we used a FWHM 
of 0.024~$\mu$m for the in\textendash plane C\textendash H stretches around 3.3~$\mu$m, of 0.2~$\mu$m for 
bands between 6 and 15~$\mu$m, of 0.3~$\mu$m for bands between 15 and 25~$\mu$m, of 
0.5~$\mu$m for bands between 25 and 50~$\mu$m and of 0.9 for bands longwards of 
50~$\mu$m. These values are just an educated guess: many factors concur in 
determining the band shape and width for IR fluorescence bands of PAHs, such 
as lifetime broadening, anharmonicity, rotational structure. For a discussion
of these effects see e.~g. \citet{pec02}. Such a detailed treatment 
would only be feasible, in principle, for those specific bands 
for which experimental measurements at different temperatures are available,
i.~e. not for the far\textendash IR ones which appear to show the strongest diagnostic
capability. Moreover it would be beyond the scope of the present work, 
since the precise band shape does not greatly affect its detectability. 

\begin{figure}
\begin{center}

\includegraphics{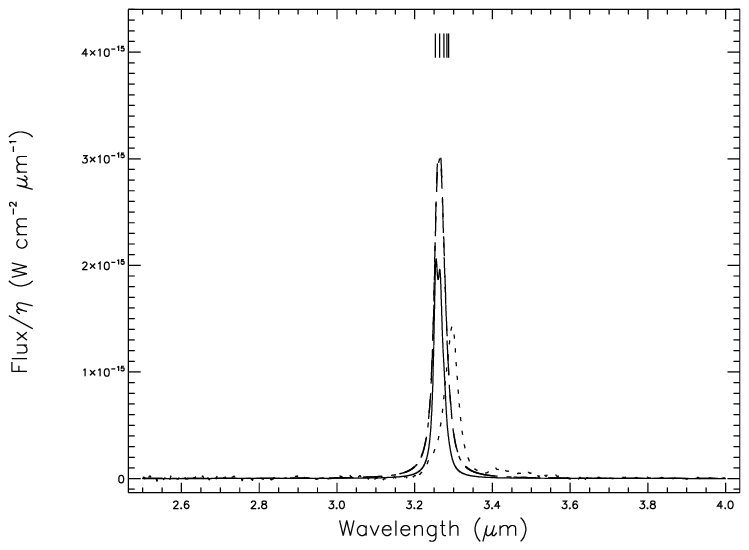}
\end{center}
\caption{Comparison between the estimated IR emission spectrum of pyrene 
(C$_{16}$H$_{10}$) and an ISO spectrum of the RR in the wavelength range 
2.5\textendash4.0~$\mu$m, including detailed band\textendash shape modelling. Calculated spectra, 
under different assumptions (see text for details), are drawn in dashed 
(calculated from the fluxes in second column of Table~\ref{pyrene_4-31G}), 
dash\textendash dotted (third column) and continuous (fourth column) lines, the 
continuum\textendash subtracted ISO spectrum is shown as a dotted line. The central 
positions of fundamental pyrene bands are marked by ticks.}
\label{pyrene_anharm}
\end{figure}

\begin{figure}
\begin{center}
\includegraphics{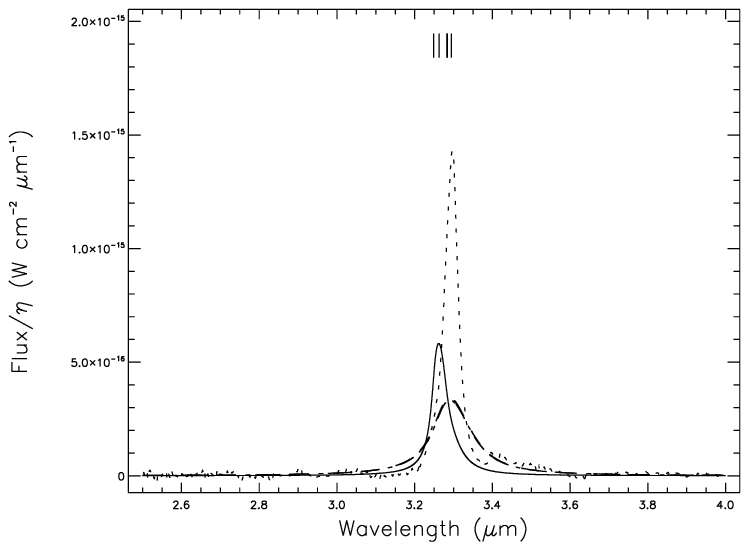}
\end{center}
\caption{Same as Fig.~\ref{pyrene_anharm} for anthracene 
(C$_{14}$H$_{10}$).}
\label{anthracene_anharm}
\end{figure}

\begin{figure}
\begin{center}
\includegraphics{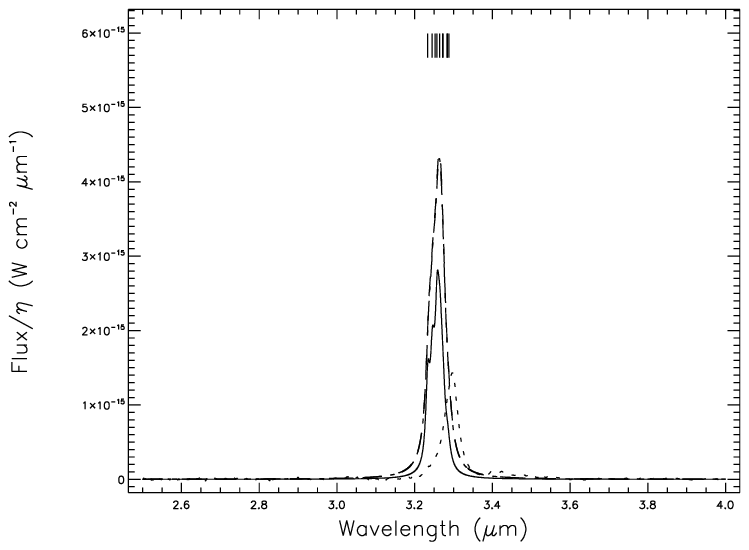}
\end{center}
\caption{Same as Fig.~\ref{pyrene_anharm} for phenanthrene (C$_{14}$H$_{10}$).}
\label{phenanthrene_anharm}
\end{figure}

We just remark that, on 
general grounds, bands which are mostly emitted when molecules are more 
highly excited are expected to be slightly redshifted due to anharmonicity
\citep{job95b,pec02}.
As an example, we performed a detailed band shape calculation for the 
in\textendash plane C\textendash H stretch of pyrene, using the experimental data from 
\citet{job95b}. Figure~\ref{pyrene_anharm} shows the result, which can be 
compared with Fig.~\ref{pyrene_3.3} to see the impact of anharmonicity. 
Since the required experimental  data are not currently available for 
phenanthrene and anthracene, we assumed for them the same anharmonicity 
parameters of pyrene. The resulting band profiles are shown in 
Figs.~\ref{anthracene_anharm} and \ref{phenanthrene_anharm}.

Considering the uncertainties in the calculated fluxes, anthracene appears 
to be compatible with all the ISO spectra, regardless of whether we use
theoretical DFT band positions or the available gas phase positions measured 
by \citet{can97} for the comparison with observations. The far\textendash IR band at
$\sim$110~$\mu$m might easily be hidden in the noise in the automatically reduced
spectra available in the online ISO database. If anthracene 
alone were responsible for the whole observed BL emission, about two thirds 
of the total observed flux in the 3.3~$\mu$m band would be due to it as well.
The slight position mismatch seen in Fig.~\ref{anthracene_3.3} can be easily 
due to anharmonicity, as shown in Fig.~ \ref{anthracene_anharm}. The band in 
the latter figure appears to be slightly too broad, but since we used the 
anharmonicity parameters of another molecule this is not definitive.

Phenanthrene, on the other hand, can possibly account at most for a small
fraction of the observed BL: its rather strong \emph{quartet} out\textendash of\textendash plane
C\textendash H bend at 13.6~$\mu$m and its skeletal bending mode at $\sim$100~$\mu$m are 
not apparently detected in the ISO spectra. The C\textendash H bend 
was measured in vapour at $\sim$770~K to be at $\sim$13.7~$\mu$m (unpublished data 
from the measurements of \citet{job92a} and \citet{job94,job95b}).
Moreover, if phenanthrene alone were responsible for the BL, it ought to 
produce more than twice the total observed flux in the 3.3~$\mu$m 
band. Even allowing for the fact that DFT calculations at the 4\textendash31G 
level are known to somewhat overestimate the IR\textendash activity of C\textendash H modes 
\citep{lan96,hud01}, this is still unrealistic. Finally, to account for the 
position mismatch between the predicted and observed band positions would
require an anharmonicity band shift about an order of magnitude larger
than the one obtained using the parameters of pyrene. Summing up, we 
conclude that fluorescence by neutral phenanthrene might contribute at 
most a small fraction of the observed BL.

The case of pyrene is similar to that of phenanthrene, although slightly 
less clear\textendash cut: it should also show a band at $\sim$100~$\mu$m which is 
undetected. This is one of the very few PAHs for which far\textendash IR 
gas\textendash phase measurements are available \citep{zha96}, and the experimental 
band origin of this fundamental vibration lies at $\sim$105.3~$\mu$m, 
unfortunately in the most noisy part of the ISO LWS spectrum available 
online. The expected flux in this band would be sufficiently quenched by a 
high photon absorption rate to make it compatible with the observed spectrum, 
if pyrene were mainly concentrated in the lobes and relatively close to 
the central source. The expected out\textendash of\textendash plane C\textendash H bend at $\sim$11.8~$\mu$m 
is a stronger constraint. This band was measured in vapour at 
$\sim$600~K to be at $\sim$11.9~$\mu$m \citep{job94,job95b}, and, as mentioned in the
discussion for phenanthrene, DFT calculations at the 4\textendash31G level are known 
to somewhat overestimate the IR\textendash activity of C\textendash H modes. Even allowing for
all of these uncertainties, the expected band at $\sim$11.8~$\mu$m is at least a 
factor of $\sim$5 too strong to be compatible with the observed spectrum.
As to the in\textendash plane C\textendash H stretch, if pyrene alone were responsible for the 
BL, it would produce all of the total observed flux in the 
3.3~$\mu$m band. However, the band position we obtain even with a detailed
band profile modelling which takes into account anharmonicity appears to
be shifted bluewards of the observed peak. Summing up, we conclude that 
neutral pyrene might contribute at most $\sim$20\% of the observed BL.

\section{Discussion and conclusions}

\citet{vij04} and \citet{vij05} put rather stringent limits on the plausible 
neutral PAHs as candidates for the BL in the RR: they restricted the range 
of possible candidates to molecules including from 3 to 4 aromatic rings. 
Among them, the only ones showing fluorescence spectral profiles compatible 
with the observed BL seem to be anthracene, phenanthrene and pyrene. 

Upon a quick examination of the browsable online data from the ISO spectral
database and comparison with our estimated IR fluxes, some 
bands stand out as the most promising diagnostics, due to their relatively 
high estimated intensities compared with ISO sensitivity: the out of plane 
C\textendash H bends and the skeletal bending modes between $\sim$100 and $\sim$110~$\mu$m. 
In particular, even a cursory comparison between the estimated IR fluxes for 
phenanthrene and the available ISO SWS and LWS spectra show that its 
contribution to the production of the BL must be very small: indeed, 
fluorescence from phenanthrene would be so inefficient both in the lobes 
and in the halo of the RR that huge column densities would be needed to 
yield a significant contribution to the BL, and they would in turn also 
yield IR band intensities incompatible with observations.
Phenanthrene and pyrene would produce prominent phosphorescence
bands too, which have not been observed in the RR thus far \citep{wit05}.  
All these elements, taken together, mean that only anthracene, among the three 
molecules considered, appears compatible with the observations.

Anthracene, if responsible for the BL, would also produce a quite substantial 
fraction of the observed flux in the in\textendash plane C\textendash H stretch bands. If it were 
to produce the observed BL, it would at the same time produce about half of 
the total observed flux at 3.3~$\mu$m. This is in good agreement with the 
observed spatial correlation between the latter band and the BL \citep{vij05}.

However, we should also consider the bands which are \emph{absent} 
from the calculated IR emission spectra: with the exception of the in\textendash plane 
C\textendash H stretch bands at 3.3~$\mu$m, \emph{all} of the other classical aromatic 
bands, which are strong in the ISO data, are almost negligible in the 
calculated spectra for the three neutral PAHs considered here. If anthracene
were responsible for the BL, then it would almost completely account for 
the observed flux in the 3.3~$\mu$m band; it follows that in this case the 
remaining aromatic bands must be produced by different carriers, which in 
turn must not contribute much flux to the 3.3~$\mu$m band. In particular, 
any mixture of neutral PAHs producing the observed flux in the out of plane 
C\textendash H bends at $\sim$11.3~$\mu$m would also necessarily produce a substantial 
contribution to the 3.3~$\mu$m band, unless only large species, with more than 
$\sim$40 C atoms, are present. A possible solution may be to consider PAH 
cations, whose in\textendash plane C\textendash H stretch bands are well\textendash known to be much less 
intense with respect to the same bands in their parent neutrals 
\citep[see e.~g. ][]{lan96,all99}. This might account for the other strong 
bands observed in the RR, including the 11.3~$\mu$m band, which is not as 
reduced by ionisation.

This is consistent with observations by \citet{bre03}, which show the
3.3 and 11.3~$\mu$m bands to have very different spatial distributions, and
would imply a scenario in which anthracene dominates the population
of small neutral PAHs, producing both the 3.3~$\mu$m band and the BL 
and PAH cations produce the other aromatic bands, possibly with some 
contribution by large neutral PAHs for the 11.3~$\mu$m band. Some detailed 
modelling would be useful to determine whether such a scenario can be 
realistic and compatible with the photo\textendash chemical evolution of PAHs 
in the RR environment.

The present data provide a powerful, quantitative diagnostic which may be used 
as a definitive cross\textendash check for the hypothesis that small, neutral PAHs are
responsible for the BL observed in the RR. Firm conclusions will require a 
thorough and careful review of available ISO spectroscopic data. In particular,
all available LWS spectra in the 90\textendash120~$\mu$m wavelength range ought to be
appropriately reduced and stacked, to maximise the S/N ratio.
The expected IR fluxes ought to be well within reach with the sensitivity 
of ISO observations, and will definitely be observable with forthcoming 
Herschel observations, which will go two orders of magnitude deeper 
(e.~g. with the PACS instrument in the spectral range 57\textendash210~$\mu$m, see the 
official PACS web page at \texttt{http://pacs.ster.kuleuven.ac.be}).

The accuracy of the IR emission spectra calculated in this paper would
be greatly improved by a better understanding of the spatial distribution
of PAH emission in the RR and, consequently, a better knowledge of the RF 
illuminating them. Indeed, the best available model of the RR \citep{men02}
does not explicitly consider PAHs. Both more observational work, to better
sample the spatial distribution of the BL as well as that of the IR emission, 
and more modelling are called for. In particular,  observations of the BL with 
a much higher spatial resolution and sampling of the RR than those by 
\citet{vij05} would be needed: this would enable us to adequately assess the 
dilution effect due to the large ISO SWS/LWS apertures without recurring to 
the extrapolations we used in Sect.~\ref{tps}.

We emphasize that experimental measurements of the precise position 
and intensities of \emph{all} IR\textendash active bands of anthracene, phenanthrene and
pyrene in astrophysically relevant conditions would be very useful and 
reduce modelling errors: while the spectral range corresponding to 
``classical'' AIBs has been well explored \citep[see e.~g.][]{job92a,szc93,
job94,job95b,mou96,coo96,hud94,hud95,hud98,all99,kim02,oom03}, gas\textendash phase 
measurements of far\textendash IR bands of PAHs are sorely lacking to date. Last but not 
least, gas\textendash phase measurements of the phosphorescence spectra of phenanthrene, 
pyrene and possibly other PAHs, which are also lacking to date, would provide 
yet another independent constraint amenable to direct observational 
verification of their presence in space.

\begin{acknowledgements}
G.~Malloci acknowledges the financial support by INAF\textendash Osservatorio
Astronomico di Cagliari. We gratefully thank Philippe Br\'echignac for
kindly making his unpublished data available for this work and Adolf Witt for
making available the original results published in \citet{vij04} and for his 
useful comments. We are thankful to the authors of \textsc{Octopus} 
for making their code available under a free license. We acknowledge the 
High Performance Computational Chemistry Group 
for using their code: ``NWChem, A Computational Chemistry Package for Parallel
Computers, version 4.6'' (2004), Pacific Northwest National Laboratory, 
Richland, Washington 99352\textendash0999, USA. Part of the calculations used here
were performed at the italian CINECA supercomputing facility.
\end{acknowledgements}

\bibliographystyle{aa}
\bibliography{biblio}

\end{document}